\documentclass{article}

\usepackage[main, final]{neurips_2025}

\usepackage[utf8]{inputenc} 
\usepackage{pdfpages}
\usepackage[T1]{fontenc}    
\usepackage{hyperref}       
\usepackage{url}            
\usepackage{booktabs}       
\usepackage{amsfonts}       
\usepackage{nicefrac}       
\usepackage{microtype}      
\usepackage{xcolor}         
\usepackage{graphicx}

\title{AgenticAD: A Specialized Multi-Agent System Framework for Holistic Alzheimer's Disease Management}

\author{%
  Adib Bazgir \\
  Department of Mechanical and Aerospace Engineering\\
  University of Missouri-Columbia\\
  Columbia, MO 65211 \\
  \And
  Amir Habibdoust \\
  Institute for Data Science and Informatics \\
  University of Missouri-Columbia\\
  Columbia, MO 65211 \\
  \AND
  Xing Song \\
  Department of Biomedical Informatics, Biostatistics, and Medical Epidemiology \\
  University of Missouri-Columbia\\
  Columbia, MO 65211 \\
  \And
  Yuwen Zhang\thanks{Corresponding to: zhangyu@missouri.edu and xsm7f@missouri.edu}\\
  Department of Mechanical and Aerospace Engineering\\
  University of Missouri-Columbia\\
  Columbia, MO 65211 \\
}

\begin{document}

\maketitle

\begin{abstract}
Alzheimer's disease (AD) presents a complex, multifaceted challenge to patients, caregivers, and the healthcare system, necessitating integrated and dynamic support solutions. While artificial intelligence (AI) offers promising avenues for intervention, current applications are often siloed, addressing singular aspects of the disease such as diagnostics or caregiver support without systemic integration. This paper proposes a novel methodological framework for a comprehensive, multi-agent system (MAS) designed for holistic Alzheimer's disease management. The objective is to detail the architecture of a collaborative ecosystem of specialized AI agents, each engineered to address a distinct challenge in the AD care continuum, from caregiver support and multimodal data analysis to automated research and clinical data interpretation. The proposed framework is composed of eight specialized, interoperable agents. These agents are categorized by function: (1) Caregiver and Patient Support, (2) Data Analysis and Research, and (3) Advanced Multimodal Workflows. The methodology details the technical architecture of each agent, leveraging a suite of advanced technologies including large language models (LLMs) such as GPT-4o and Gemini, multi-agent orchestration frameworks (AutoGen, OpenAI Agents SDK, Agno), Retrieval-Augmented Generation (RAG) for evidence-grounded responses, and specialized tools for web scraping (Firecrawl, ScrapegraphAI), multimodal data processing (vision, audio), and in-memory database querying (DuckDB). This paper presents a detailed architectural blueprint for an integrated AI ecosystem for AD care. By moving beyond single-purpose tools to a collaborative, multi-agent paradigm, this framework establishes a foundation for developing more adaptive, personalized, and proactive solutions. This methodological approach aims to pave the way for future systems capable of synthesizing diverse data streams to improve patient outcomes and reduce caregiver burden.
\end{abstract}

\section{Introduction}
\label{headings}
Alzheimer's disease (AD) represents a significant and escalating global health crisis, characterized by a progressive decline in cognitive function that profoundly impacts patients, their families, and healthcare systems [1]. The management of AD is not a singular clinical challenge but a complex, long-term process involving medical treatment, cognitive support, behavioral management, and immense caregiver burden [2]. Caregivers must navigate a vast and evolving landscape of information regarding symptoms, safety protocols, and daily care strategies, often with limited professional support [3]. This complexity necessitates technological solutions that are not only powerful but also integrated, adaptive, and holistic.
The advent of artificial intelligence (AI), particularly advancements in machine learning (ML), deep learning (DL), and large language models (LLMs), has introduced transformative potential in medicine [4]. In the context of dementia, AI applications are emerging across the entire care continuum, including early diagnosis through neuroimaging analysis, prediction of disease progression, drug discovery, and patient monitoring [5]. LLMs, for instance, are being leveraged to provide companionship, summarize clinical records, and offer therapeutic assistance [6]. However, many of these powerful tools are developed and deployed in isolation, creating information silos that limit their overall effectiveness. A diagnostic model that analyzes an MRI scan does not communicate with a caregiver support chatbot, and a research tool that scrapes literature does not integrate with a system monitoring a patient's daily activities. This fragmentation mirrors the often-disjointed nature of healthcare itself and represents a critical barrier to providing truly comprehensive, data-driven care [7].

To overcome these limitations, a paradigm shift from monolithic, single-purpose applications to integrated, collaborative systems is required. Multi-Agent Systems (MAS) offer a compelling architectural solution. A MAS is composed of multiple autonomous, intelligent agents that communicate and coordinate to solve problems beyond the scope of any single agent [7]. This distributed, modular approach is exceptionally well-suited to the multifaceted nature of AD, allowing for the creation of an ecosystem of specialized agents that can work in concert. Such a system can enhance information sharing, enable adaptive and personalized care, and optimize complex clinical and caregiving workflows.
This paper presents the methodological framework and detailed architecture of a novel, comprehensive MAS designed for Alzheimer's management. We detail the design and implementation of eight core, specialized agents that constitute this proposed ecosystem. These agents are engineered to perform distinct but complementary functions, spanning direct caregiver support, automated research, structured and unstructured data analysis, and multimodal information processing. By elucidating the technical underpinnings of each component, we provide a blueprint for an integrated system that moves beyond isolated AI tools toward a truly holistic and collaborative platform for dementia care.

\section{Methodology: A Multi-Agent Framework for Alzheimer's Care}
The proposed framework, as shown in Figure 1, is an integrated ecosystem of eight specialized AI agents. These agents are designed to be modular and interoperable, each addressing a specific need within the Alzheimer's disease care and research landscape. They are grouped into three functional categories: 
\begin{enumerate}
    \item Caregiver and Patient Support Agents
    \item Data Analysis and Research Agents
    \item Advanced Multimodal and Workflow Agents
\end{enumerate}

\subsection{Caregiver and Patient Support Agents}

These agents are designed for direct interaction with caregivers and patients, providing evidence-based information, personalized support, and structured guidance.

\subsubsection{Alzheimer’s Support Agent}
The objective is to generate a comprehensive and personalized dementia care plan based on detailed user inputs regarding a patient's condition and environment. This agent is implemented as a multi-agent swarm using the Autogen framework. It consists of three distinct sub-agents: an Assessment Agent, a Care Plan Agent, and a Follow-up Agent, all powered by the GPT-4o LLM. The user interface is built with Streamlit. The system collects structured data from the user (e.g., age, symptoms, living situation, safety concerns). The Assessment Agent first synthesizes this information into a clinical summary. This summary is then passed to the Care Plan Agent, which proposes practical daily plans, safety mitigations, and clinical next steps. Finally, the Follow-up Agent receives the preceding outputs and generates a long-term monitoring plan, including tracking templates and escalation criteria. The agents coordinate through a shared context and explicit handoffs, ensuring a cohesive, multi-part report.

\subsubsection{Alzheimer’s PDF Assistant Agent}
For this agent, the objective is to provide a conversational interface that allows users to ask questions and receive evidence-grounded answers from a curated knowledge base of PDF documents. This agent employs a Retrieval-Augmented Generation (RAG) architecture. It is built using the embedchain library, which orchestrates an OpenAI embedding model, a Chroma vector database for storage, and an OpenAI LLM for response generation. The UI is developed with Streamlit. Users upload one or more PDF files (e.g., clinical guidelines, caregiver manuals), which are chunked, vectorized, and stored in a session-specific Chroma database. When a user poses a query, the system performs a similarity search to retrieve the most relevant text chunks from the document corpus. These chunks are then prepended to the user's prompt as context for the LLM, which generates an answer grounded in the provided sources, thereby mitigating hallucination and ensuring factual accuracy.

\begin{figure*}
  \centering
  \includegraphics[width=1\textwidth]{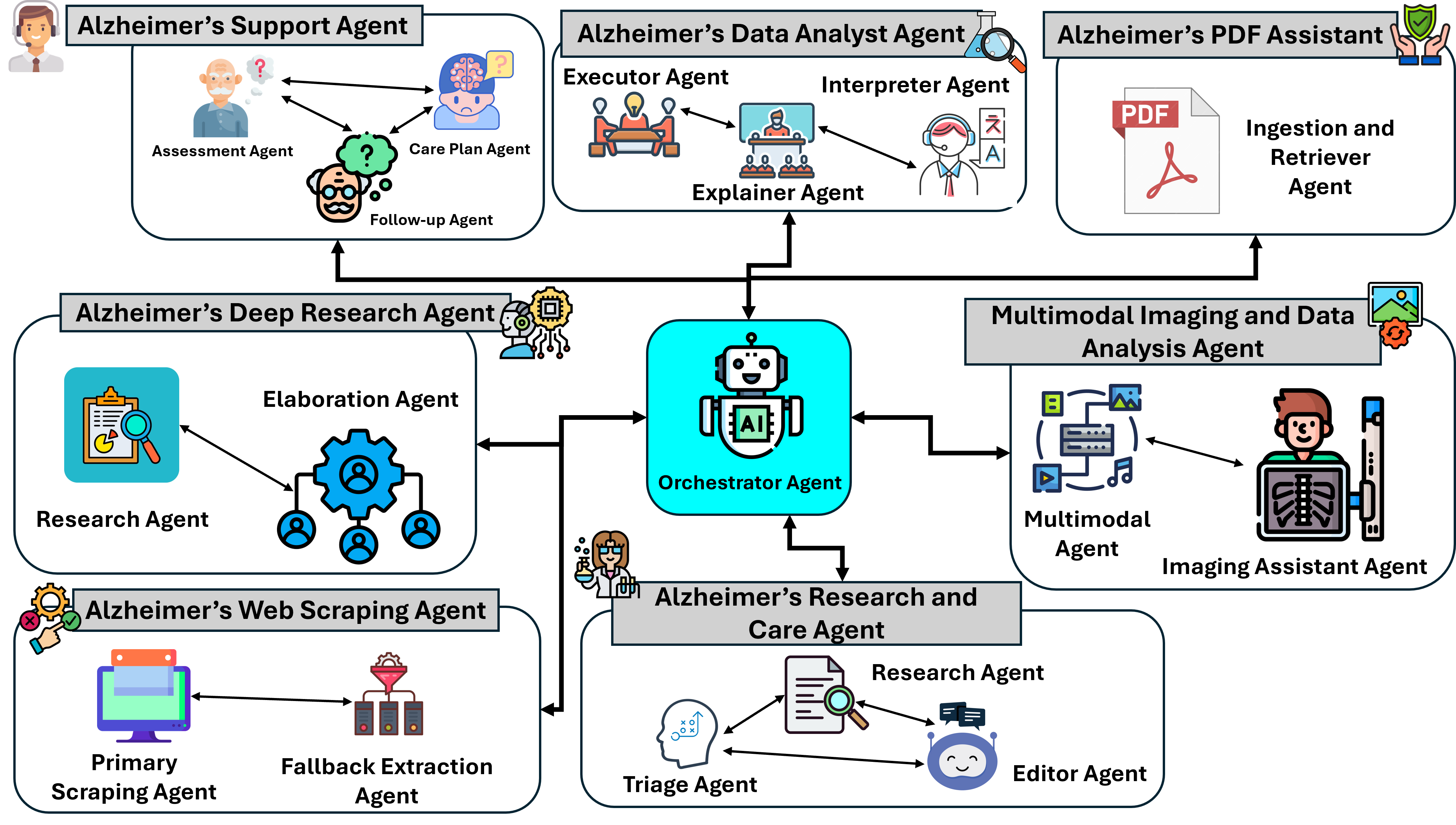}
  \caption{High-Level Architecture of the AgenticAD AI Ecosystem.}
  \label{fig:agentic system}
\end{figure*}

\subsection{Data Analysis and Research Agents}

This group of agents is engineered to automate the collection, extraction, and analysis of both unstructured and structured data from web sources and user-provided datasets.

\subsubsection{Alzheimer’s Deep Research Agent}
The objective is to perform deep, automated web research on a specified topic and produce a structured, multi-section report. This system uses a two-stage workflow with AutoGen SwarmAgents and the Firecrawl web scraping service. The backend is powered by the GPT-4o LLM, with a Streamlit interface. The workflow begins with a Research Agent that takes a user's topic and uses the Firecrawl tool to perform an in-depth crawl of web sources. It then synthesizes the scraped information into an initial structured report. This report is passed to a second Elaboration Agent, which enhances the content by adding deeper explanations, examples, and future outlooks, resulting in a more comprehensive final document.

\subsubsection{Alzheimer’s Web Scraping Agent}
The objective for this agent is to extract specific, structured information from a single webpage based on a natural language prompt. This agent utilizes the ScrapegraphAI library, which leverages an LLM (GPT-4o-mini or GPT-4o) to parse webpage content. It includes a robust fallback mechanism that uses the requests and BeautifulSoup libraries for basic text extraction and a direct call to the OpenAI API with a user-defined JSON schema for structured data extraction if the primary method fails. The user provides a URL and a prompt (e.g., ``Extract the author, date, and key findings''). ScrapegraphAI attempts to fulfill the request. If it encounters a parsing error, the fallback system fetches the page's raw text, truncates it, and sends it to the LLM with explicit instructions to populate a predefined JSON object, ensuring a high likelihood of successful data extraction.

\subsubsection{Alzheimer’s Data Analyst Agent}
In this agent role, the objective is to enable natural language querying of structured datasets (CSV, XLSX) by translating user questions into executable SQL queries. This agent leverages the Agno agent framework with an OpenAI LLM (GPT-4o-mini) as its reasoning engine. It integrates with a DuckDB in-memory analytical database for high-performance SQL execution and uses the Pandas library for data preprocessing. A user uploads a de-identified dataset. The data is loaded into a DuckDB instance and exposed as a table named uploaded\_data. The user asks a question in natural language (e.g., ``What is the monthly trend of agitation incidents?''). The Agno agent, guided by a system prompt that specifies the table schema and instructs it to generate DuckDB-compatible SQL, produces a SQL query. The application extracts this query, executes it against the database, and displays the resulting data frame to the user.

\subsection{Advanced Multimodal and Workflow Agents}

This category comprises agents designed to handle complex, multi-step workflows and process diverse, non-textual data modalities.

\subsubsection{Alzheimer’s Research and Care Agent}
This agent works to orchestrate a sophisticated, multi-agent research and report-generation workflow to produce a comprehensive, caregiver-friendly brief on an Alzheimer's-related topic. This system is built on the OpenAI Agents SDK, which allows for the definition of specialized agents with distinct instructions and tools that can hand off tasks to one another. It features a Triage Agent, a Research Agent, and an Editor Agent, using GPT-4o-mini. Pydantic models are used to enforce structured outputs. The workflow is initiated by the Triage Agent, which receives the user's topic and generates a structured research plan, including specific search queries biased toward reputable medical sources. This plan is handed off to the Research Agent, which executes the queries using a built-in web search tool and synthesizes the findings. Finally, the collected information is passed to the Editor Agent, which writes a detailed, multi-section report in Markdown format, adhering to a predefined structure suitable for caregivers.

\subsubsection{Alzheimer’s Multimodal Agent}
This agent's objective is to analyze and synthesize information from multiple data modalities, including images, audio, and video, in conjunction with web search results to generate a holistic brief. This agent is built using the Agno framework and is powered by a multimodal foundation model, Google Gemini. It is equipped with a DuckDuckGo web search tool to augment its analysis with external information. The user can upload an image, audio file, and/or video file and provide a text-based prompt. The agent processes all inputs simultaneously. The system prompt is dynamically constructed to guide the Gemini model, specifying the target audience, key focus areas, and a strong preference for reputable medical sources when using its web search tool. The agent's output is a unified, context-aware response that integrates insights from all provided media.

\subsubsection{Alzheimer’s Imaging Assistant Agent}
The aim of this agent is a specialized variant of the Multimodal Agent focused on the educational analysis of brain imaging files (e.g., PNG, JPG, DICOM). This agent uses the same technical stack as Agent 5 (Agno, Google Gemini, DuckDuckGo tools) but is equipped with a highly specialized and safety-oriented system prompt. The agent is designed to receive a brain imaging file and provide a structured, non-diagnostic analysis. Its prompt explicitly instructs it to follow a multi-step process: identify the imaging modality, describe key observations in qualitative terms, discuss potential dementia-related considerations without assigning a diagnosis, list urgent red-flag symptoms, and provide a caregiver-friendly explanation. It uses its web search tool to find and cite reputable resources, reinforcing its role as an educational, not a clinical, tool.

\section{Results and Discussion}

This section presents the operational results of each of the eight specialized agents, followed by a discussion of their performance, significance, and implications within the broader context of the proposed multi-agent framework for Alzheimer's care.

\subsection{Alzheimer’s Support Agent}

The Alzheimer's Support Agent was tested across three distinct user personas: ``Clinician / Staff,'' ``Caregiver / Family Member,'' and ``Person with Memory Concerns.'' In each scenario, the agent successfully generated a structured, multi-part care plan tailored to the specific inputs provided. The output consistently comprised three main sections, generated sequentially by the agent swarm: an Assessment Synthesis, a Safety-First Daily Care Plan, and a Follow-up and Monitoring Plan. The agent demonstrated a high degree of personalization. For instance, when the user was a Caregiver whose primary goal was establishing a ``routine'' for a 76-year-old individual with paranoia and aggression, the generated plan emphasized monthly medication reviews for donepezil and tracking of agitation episodes. In contrast, for a Clinician user focused on ``safety'' for a 76-year-old living alone with poor judgment and frequent falls, the plan was more clinically framed, highlighting medication adherence for memantine and sensitive prompts for discussing driving capabilities. A core component of the agent's output is the detailed Follow-up Plan. This section was consistently structured across all test cases, providing actionable guidance under five key headings:
\begin{enumerate}
    \item \textbf{Check-in Cadence:} Recommending weekly notes and monthly medication/behavior reviews.
    \item \textbf{Tracking Template:} Providing a checklist of key metrics to monitor, such as sleep hours, agitation episodes, falls, meals, and medication adherence.
    \item \textbf{Escalation Criteria:} Clearly defining when to contact a clinician versus seeking urgent care/ER based on specific symptoms like acute confusion, new neurological deficits, or injuries from falls.
    \item \textbf{Care Progression Planning:} Offering sensitive, conversation-starting prompts for difficult topics like driving safety and legal/financial planning.
    \item \textbf{Resource Refresher:} Directing users to established support organizations like the Alzheimer’s Association and local Area Agencies on Aging.
\end{enumerate}

The results indicate that the multi-agent swarm architecture is effective at breaking down the complex task of care planning into a logical, sequential workflow, producing a cohesive and contextually relevant output for different end-users. The performance of the Alzheimer's Support Agent validates the utility of a multi-agent approach for complex, user-facing healthcare tasks. The agent directly addresses a critical gap in dementia care: the need for accessible, structured, and personalized information for caregivers who face immense emotional, physical, and financial strain. The caregiving role is multifaceted, requiring knowledge of the disease, management of medical and financial duties, and the provision of constant emotional support. By generating a comprehensive plan, the agent functions as a cognitive offloading tool, reducing the mental burden on caregivers and providing a clear framework for action, which is a crucial factor in preventing caregiver burnout.

Architecturally, this agent serves as a microcosm of the larger MAS concept. The use of three specialized sub-agents (Assessment, Care Plan, Follow-up) demonstrates the principle of distributed problem-solving, where each component handles a distinct sub-task before handing off its output to the next specialist. This modularity is a key advantage of MAS, allowing for more focused and maintainable components compared to a single, monolithic model attempting to perform the entire sequence of tasks. However, the results also highlight a key limitation inherent in a system reliant on self-reported data. The agent's output is entirely contingent on the accuracy and completeness of the information provided by the user. It does not have access to, nor can it verify its inputs against, formal electronic health records (EHRs) or other clinical data sources. This underscores the necessity for future integration-level enhancements, where such support agents can securely interact with verified patient data to provide more robust and clinically aligned recommendations. While this agent provides crucial supportive guidance based on user input, the next agents in the framework are designed to address the challenge of acquiring and analyzing data from external sources.

\subsection{Alzheimer’s Deep Research Agent}

The Alzheimer’s Deep Research Agent was evaluated on its ability to conduct automated web research and generate structured reports. In a successful test run, the agent was tasked with researching the ``Latest development of large language models for Alzheimer's disease.'' The agent executed its two-stage workflow as designed. First, it produced a concise INITIAL Report that provided a high-level summary of the topic. Subsequently, the Elaboration Agent generated a more detailed ENHANCED Report. This enhanced version was significantly more structured, containing specific sections such as Early Detection and Diagnosis, Drug Discovery and Development, and Practical Implications for Stakeholders. It also included specific examples, such as a case study referencing a 2023 publication, and a forward-looking analysis of trends like personalized medicine and integration with wearable technology. However, testing also revealed a critical failure mode. When tasked with researching ``The application of multi-agent LLMs for Alzheimer's disease science,'' the agent failed to execute. The system returned an error message stating, ``It appears there's a persistent issue with accessing the research tool needed to gather information on the topic,'' indicating that a failure in the external Firecrawl API prevented the entire workflow from completing. This demonstrates that the agent's functionality is entirely dependent on the availability and correct configuration of its external tool.

The successful execution of this agent demonstrates the potential for AI to automate and accelerate the process of scientific literature and information review, a foundational task in both clinical practice and research. The two-stage architecture, moving from a brief initial summary to an in-depth enhanced report, mirrors the human process of information gathering and synthesis. This approach allows for rapid topic assessment followed by a more comprehensive analysis, making it a potentially powerful tool for researchers, clinicians, and students seeking to stay abreast of rapid developments in fields like Alzheimer's research. The quality of the enhanced report, with its structured sections and forward-looking analysis, suggests that such agents can move beyond simple information retrieval to perform valuable synthesis and summarization. The failure case, however, is equally instructive. It highlights a critical vulnerability of agent-based systems that rely on external tools and APIs: dependency risk. The agent's inability to function without a connection to the Firecrawl service underscores that the robustness of the entire system is limited by the reliability of its individual components and their connections. This is a crucial consideration for the design of any real-world MAS, which will inevitably need to integrate with numerous external data sources, from public APIs to hospital EHR systems. Future iterations of such agents would require more sophisticated error handling and fallback mechanisms, such as the ability to query alternative data sources or notify a human operator when a critical tool is unavailable. Within the proposed MAS framework, this agent serves as a vital knowledge acquisition component. Its function is to transform the vast, unstructured information of the open web into a structured, curated document. This output could then serve as a foundational knowledge source for other agents. This illustrates the synergistic potential of the MAS, where one agent's primary output becomes another's primary input, creating a value chain of information processing and refinement.

\subsection{Alzheimer’s Web Scraping Agent}

The Alzheimer’s Web Scraping Agent was tested on its ability to extract targeted information from reputable health websites. In two separate tests, the agent was tasked with extracting information on the causations of Alzheimer's disease from a Mayo Clinic webpage and information on treatment from a Cleveland Clinic webpage. In both instances, the primary scraping method, SmartScraperGraph, failed to execute. The system returned an identical error message related to a missing Playwright executable, indicating a dependency configuration issue rather than a failure of the scraping logic itself. Critically, the agent's built-in fallback mechanism was successfully triggered in both cases. The fallback system completed the task and produced a Strict JSON Fallback Output. This output was a structured JSON object containing a summary and an array of key\_points. The extracted content was accurate and relevant to the queries. For the Mayo Clinic URL, the agent correctly identified that Alzheimer's is caused by a combination of genetic, lifestyle, and environmental factors and is characterized by amyloid plaques and tau tangles. For the Cleveland Clinic URL, the agent correctly summarized treatment options, including cholinesterase inhibitors, NMDA antagonists, and monoclonal antibodies like lecanemab and donanemab.

The results for this agent highlight a crucial principle in the design of robust agent-based systems which is fault tolerance. While the primary, more sophisticated scraping tool failed due to an environmental dependency issue, a common problem in real-world software deployment, the agent's architectural resilience allowed it to complete its mission successfully. The two-tiered design, which combines a complex primary tool with a simpler, more robust fallback, ensures a high degree of reliability. This agent's function within the broader MAS is that of a precision data extractor. Unlike Alzheimer's Deep research agent, which performs broad, topic-based research, this agent is designed for surgical extraction of specific facts from a known source. Its ability to convert unstructured HTML content into a structured JSON format is a fundamental data ingestion and normalization task. This structured output is machine-readable and can be seamlessly consumed by other agents in the ecosystem. For example, the key points extracted by this agent could be used to populate a section in a report being generated by next agents, or they could serve as a concise, factual answer to a specific query posed to Alzheimer's support agent. The failure of SmartScraperGraph due to a missing dependency serves as a practical illustration of the engineering challenges in building and maintaining complex AI systems. The successful operation of the fallback mechanism, which relies on more fundamental libraries (requests and BeautifulSoup) and a direct, schema-guided LLM call, demonstrates a best practice in system design. It prioritizes task completion over methodological purity, ensuring that the user receives a valuable result even when the preferred pathway is unavailable. This emphasis on resilience is paramount for any AI system intended for deployment in critical domains like healthcare, where system uptime and reliability are non-negotiable.

\subsection{Alzheimer’s Research and Care Agent}
The Alzheimer’s Research and Care Agent was evaluated on its ability to execute a complete, end-to-end research and report generation workflow. The agent was tested on three distinct, caregiver-relevant topics: Managing agitation and sundowning at home, What evaluations are typical when memory loss is first noticed?, and Non-pharmacologic approaches for sleep problems in dementia. In all three test cases, the agent successfully completed its three-stage \emph{Triage $\rightarrow$ Research $\rightarrow$ Edit} workflow. The initial Triage Agent consistently produced a structured Research Plan that refined the user's topic and generated a list of 3--5 specific search queries and corresponding focus areas. Notably, these queries were automatically augmented with filters to prioritize reputable sources such as \texttt{nia.nih.gov}, \texttt{alz.org}, and \texttt{mayoclinic.org}, ensuring the subsequent research phase was grounded in high-quality information. Following the research phase, the final Editor Agent generated a comprehensive, long-form report for each topic. The reports were consistently over 1000 words in length and followed a predefined, caregiver-friendly structure, including sections such as \emph{Executive Summary}, \emph{Evidence Review}, \emph{Practical Care Tips}, \emph{Red-Flags and When to Seek Care}, and \emph{Resources and Helplines}. The content was detailed, well-organized, and written in accessible language appropriate for a non-clinical audience.

The performance of this agent represents a significant step forward in architectural sophistication compared to the preceding agents. It successfully demonstrates a complete, orchestrated workflow automation, a hallmark of an advanced Multi-Agent System (MAS). The agent's ability to autonomously plan, execute, and synthesize information into a polished final product showcases the potential for MAS to handle complex, multi-step knowledge work with minimal human intervention. The Triage Agent's strategic planning phase is a particularly noteworthy feature. By proactively formulating search queries that target authoritative domains, the system embeds a crucial best practice for medical information retrieval directly into its operational logic. This automated source-vetting is a powerful mechanism for enhancing the reliability and trustworthiness of the final output, a critical requirement for any AI system providing health-related information. Furthermore, the final report generated by the Editor Agent is a high-value asset for the target audience. The consistent structure and comprehensive nature of the output provide caregivers with actionable, easy-to-digest information that directly addresses their most pressing concerns. This aligns with the broader goal of using AI to empower caregivers, reduce their informational burden, and support them in their challenging roles. Within the proposed MAS framework, this agent functions as a synthesis and orchestration engine. It is capable of taking a high-level user request and managing a complex pipeline of sub-tasks to deliver a complete product. This agent could theoretically orchestrate other specialized agents in the ecosystem; for instance, it could delegate the initial web crawling to Alzheimer's deep research agent and the precision fact-extraction to Alzheimer's web scraping agent before synthesizing their outputs into its final report. This demonstrates the scalable and hierarchical potential of MAS, where a manager agent can coordinate the work of specialist agents to achieve a complex, system-level goal.

\subsection{Multimodal Imaging and Data Analysis Agents}

The multimodal agents were evaluated on their capacity to analyze and interpret complex, non-textual data, specifically brain imaging files. These agents were tested with two different brain scan images. In both cases, the agents successfully processed the visual data and generated structured, five-part analysis as dictated by their safety-oriented prompt. For the first image, the agent correctly identified the likely modality as a coronal MRI or PET scan and noted a moderate to severe asymmetry between the brain hemispheres as the primary finding. For the second image, it identified a large, well-defined area of abnormal signal intensity in the left frontal lobe. In both tests, the agent provided a list of possible, non-diagnostic considerations (e.g., neurodegenerative diseases, stroke, tumor) and a simplified, patient-friendly explanation of the findings. A key capability demonstrated was the successful integration of its internal analysis with external knowledge retrieval. After analyzing the visual data, the agent autonomously used its DuckDuckGo search tool to find relevant research context. For the second image, it explicitly searched for Large Frontal Lobe Lesions on MRI and summarized key takeaways, providing links to reputable resources like the Radiology Society of North America (RSNA) and the National Institute of Neurological Disorders and Stroke (NINDS). This demonstrates a successful fusion of computer vision-based analysis with text-based web retrieval to produce a comprehensive and contextually enriched output.

The successful performance of these multimodal agents is critical, as it addresses a fundamental requirement for any advanced clinical AI system: the ability to interpret heterogeneous, real-world medical data. Clinical decision-making is rarely based on a single data type; it requires the synthesis of information from neuroimaging, lab results, clinical notes, and patient-reported outcomes. These agents represent the sensory and perceptual layer of the proposed MAS, capable of translating complex, raw data like an MRI scan into a structured, machine-readable format that other agents can utilize. The agent's design demonstrates a strong emphasis on responsible AI principles. The heavily constrained, safety-first system prompt forces the agent to frame its output as educational and explicitly non-diagnostic, a crucial guardrail for preventing misuse and ensuring the tool supports, rather than replaces, the clinician. The inclusion of a patient-friendly explanation addresses the need for improved patient-provider communication and shared decision-making, while the automated retrieval of research context from vetted sources enhances the transparency and trustworthiness of the analysis. This capability is central to the vision of a holistic AI ecosystem. The structured analysis produced by this agent could be passed to Alzheimer's support agent to inform the creation of a more clinically aware care plan. For example, if the imaging analysis notes significant left-hemisphere involvement, the care plan could be automatically prompted to include specific monitoring for language-related symptoms. Similarly, the findings could be incorporated into a research report by Alzheimer's research and care agent. This seamless flow of information, from raw pixel data to structured analysis to actionable care planning, is the core value proposition of an integrated MAS, enabling a level of data synthesis and proactive care that is impossible with siloed, single-modality tools.

\subsection{Alzheimer’s Data Analyst Agent}

The Alzheimer’s Data Analyst Agent was evaluated on its ability to perform natural language querying on a structured and de-identified patient dataset. The agent successfully ingested the dataset, which contained 35 columns of demographic, clinical, and behavioral data. Two test queries were executed. The first query was to ``predict the average age of patients with BMI greater than 24.'' The agent correctly interpreted this request, generated the valid SQL query:
\begin{verbatim}
SELECT AVG(age) AS average_age FROM uploaded_data WHERE bmi > 24;
\end{verbatim}
and the backend database executed it to return the result of 74.8731.

The second query asked the agent to ``Use ML analysis and predict the value of Alcohol Consumption for a patient at the age of 66.'' Again, the agent translated this into a descriptive SQL query:
\begin{verbatim}
SELECT AVG(AlcoholConsumption) AS predicted_alcohol_consumption 
FROM uploaded_data WHERE Age = 66;
\end{verbatim}
which returned a value of 10.0227. In both cases, the agent provided a plain-language explanation of the query and, for the second query, included a critical caveat that ``individual consumption can vary significantly and should be interpreted with caution.''

The performance of this agent demonstrates a powerful capability for democratizing data analysis in a clinical or research context. By translating natural language questions into executable SQL, the agent effectively removes the technical barrier to entry for exploring structured datasets. This allows users without programming expertise, such as clinicians, caregivers, or patient advocates, to directly query data and derive quantitative insights. A crucial aspect of the agent's behavior is its conservative and safe interpretation of ambiguous user prompts, particularly the word ``predict.'' Instead of attempting to build a complex and potentially uninterpretable machine learning model to predict an individual outcome, a task fraught with ethical and technical challenges, the agent defaults to a transparent and statistically sound descriptive analysis (i.e., calculating a conditional average). This design choice aligns with the principle of interpretability, which is paramount in healthcare applications where ``black box'' models can erode trust and obscure the reasoning behind a decision. The agent provides a descriptive prediction for a cohort, which is a verifiable fact from the data, rather than an inferential prediction for an individual, which would be a probabilistic estimate. Within the MAS framework, this agent functions as the structured data interpreter. It is the component responsible for extracting specific, quantitative facts from large, tabular datasets like EHR exports or clinical trial results. The numerical outputs from this agent can serve as high-value inputs for other agents. For instance, a statistic generated by this Agent could be seamlessly integrated into a research report by Alzheimer's research and care agent or used by Alzheimer's support agent to provide a data-driven rationale for a specific care plan recommendation. This agent, therefore, acts as a critical bridge between raw, structured data and actionable, human-understandable knowledge.

\subsection{Alzheimer’s PDF Assistant}

The Alzheimer’s PDF Assistant was tested on its ability to ingest and answer questions based on user-uploaded scientific articles. In two separate tests, the agent was provided with a PDF and a natural language query about its contents. In the first test, the agent was given a research paper and asked to ``summarize the main results of this work.'' It successfully extracted and presented a four-point summary covering the paper's key findings on individual clinical profiles, blood-based biomarkers, and early diagnosis. In the second test, the agent was given a different paper and asked to ``list out the results of this work.'' The agent provided a more detailed, structured summary of the \emph{Ginkgo Evaluation of Memory Study (GEMS)}, including specific quantitative details about the study's duration and participant numbers, as well as key findings related to beta-amyloid levels. Crucially, this response included a citation to the source document, demonstrating its ability to maintain data provenance.

The performance of this agent provides a clear demonstration of the value of the Retrieval-Augmented Generation (RAG) architecture in a medical context. RAG is a critical technique for building trustworthy LLM applications, as it grounds the model's responses in a specific, verifiable corpus of information, thereby mitigating the risk of factual inaccuracies or hallucinations. This is particularly important in healthcare, where the accuracy and reliability of information are paramount. A systematic review and meta-analysis of RAG in biomedicine found that its implementation leads to a statistically significant increase in performance compared to baseline LLMs. Within the proposed MAS, this agent serves as the curated knowledge expert. It enables the creation of highly specialized, private knowledge bases that can be tailored to the needs of a specific user, whether a caregiver uploading support manuals or a clinician uploading the latest clinical trial data. This capability for rapid, on-demand knowledge base creation is a powerful tool for personalization. The true potential of this agent is realized through its synergy with other agents in the ecosystem. The static, long-form reports generated by the research agents can be ingested by this agent, transforming them from read-only documents into interactive, conversational resources. This creates a powerful workflow: broad, automated research is conducted, the findings are synthesized into a curated report, and that report then becomes a dynamic knowledge base that users can query in natural language. While the current implementation is highly effective for single-document question-answering, it is limited by the vanilla RAG architecture, which can struggle with complex questions requiring multi-step reasoning across multiple sources. Future work should explore iterative RAG techniques that allow the agent to perform multiple rounds of information-seeking to synthesize answers to more complex clinical queries.

\section{Conclusion}

This paper has presented a detailed architectural and methodological framework for a comprehensive Multi-Agent System (MAS) designed to address the multifaceted challenges of Alzheimer's disease. Through the individual design and evaluation of eight specialized agents, we have demonstrated the viability of applying a diverse suite of modern AI technologies, from multi-agent swarms and Retrieval-Augmented Generation (RAG) to multimodal analysis and natural language database querying, to distinct problems across the dementia care continuum. The results confirm that specialized agents can effectively execute complex, domain-specific tasks. The support agents provide personalized, evidence-grounded information that can serve as a crucial cognitive offloading tool for overburdened caregivers. The research and data analysis agents showcase the potential to automate knowledge acquisition and democratize data analysis, transforming unstructured web content and complex datasets into actionable insights. Finally, the advanced workflow and multimodal agents illustrate the capacity for sophisticated workflow orchestration and the interpretation of critical non-textual data like neuroimaging, all while operating within safety-conscious, non-diagnostic frameworks. However, the true significance of this framework lies not in the performance of any single agent, but in the emergent potential of their integration. The architectural blueprint presented here is predicated on the principle of synergy: the structured report from the Research Agent can become the verifiable knowledge base for the PDF Assistant; the quantitative analysis from the Data Analyst can inform the personalized recommendations of the Support Agent; and the imaging insights from the Multimodal Agent can trigger a proactive care planning workflow managed by the Orchestration Agent. This collaborative potential represents a paradigm shift from the current landscape of siloed AI tools to a holistic ecosystem capable of synthesizing heterogeneous data into a coherent, patient-centric model of care. This work also highlights the critical challenges that must be addressed to realize this vision. The operational dependencies and failure modes observed in agents reliant on external tools underscore the need for robust, fault-tolerant engineering. More importantly, while this paper provides the methodological foundation, the next frontier is technical integration and clinical validation. Future research must focus on developing secure, interoperable communication protocols and sophisticated orchestration layers that allow these disparate agents to collaborate effectively and safely. Concurrently, these systems must be subjected to rigorous, prospective clinical trials to validate their efficacy, safety, and real-world impact on patient and caregiver outcomes. In conclusion, the transition from single-purpose AI applications to integrated Multi-Agent Systems is a necessary and logical evolution for the field of dementia care. The eight agents detailed in this paper serve as the foundational components of such a system, each providing a critical capability. By building upon this methodological framework and tackling the challenges of integration and validation, the field can move closer to the ultimate goal of P4 medicine, a future where Alzheimer's care is Predictive, Personalized, Preventive, and Participatory, transforming the management of this devastating disease and improving the quality of life for millions.

\newpage
\section*{References}

{
\small

[1] Dobbins, N., Xiong, C., Lan, K. \& Yetisgen, M.\ (2025) Large language model-based agents for automated research reproducibility: An exploratory study in Alzheimer’s disease. {\it arXiv preprint} arXiv:2505.23852.

[2] Gao, S., Fang, A., Huang, Y., Giunchiglia, V., Noori, A., Schwarz, J.R., Ektefaie, Y., Kondic, J. \& Zitnik, M.\ (2024) Empowering biomedical discovery with AI agents. {\it Cell} {\bf 187}(22):6125--6151.

[3] Hou, W., Yang, G., Du, Y., Lau, Y., Liu, L., He, J., Long, L. \& Wang, S.\ (2025) ADAgent: LLM agent for Alzheimer’s disease analysis with collaborative coordinator. {\it arXiv preprint} arXiv:2506.11150.

[4] Gorenshtein, A., Shihada, K., Sorka, M., Aran, D. \& Shelly, S.\ (2025) LITERAS: Biomedical literature review and citation retrieval agents. {\it Computers in Biology and Medicine} {\bf 192}:110363.

[5] Li, R., Wang, X., Berlowitz, D., Mez, J., Lin, H. \& Yu, H.\ (2025) CARE-AD: A multi-agent large language model framework for Alzheimer’s disease prediction using longitudinal clinical notes. {\it NPJ Digital Medicine} {\bf 8}(1):541.

[6] Cummings, J., Feldman, H.H. \& Scheltens, P.\ (2019) The “rights” of precision drug development for Alzheimer’s disease. {\it Alzheimer’s Research \& Therapy} {\bf 11}(1):76.

[7] Khalil, R.A., Ahmad, K. \& Ali, H.\ (2025) Redefining elderly care with agentic AI: Challenges and opportunities. {\it arXiv preprint} arXiv:2507.14912.

}

\newpage
\appendix

\section{Technical Appendices and Supplementary Material}

This appendix provides the unabridged sample outputs that serve as the primary empirical evidence for the analysis presented in the "Results and Discussion" section of this paper. These documents offer a transparent and detailed view of each specialized agent's performance, allowing the reader to directly observe their architectural capabilities, the quality of their generated content, and the specific mechanisms discussed in the main text. The materials are organized to correspond with the agents evaluated in Section 3. To facilitate a clear understanding, we encourage the reader to note the following specific features within the provided outputs:

\begin{itemize}
    \item \textbf{For the Alzheimer’s Support Agent:} Observe the distinct outputs generated for the \texttt{"Clinician,"} \texttt{"Caregiver,"} and \texttt{"Person with Memory Concerns"} user personas. Pay particular attention to how the tone, terminology, and specific action items within the \texttt{"Follow-up \& Monitoring Plan"} are tailored to each audience, demonstrating the agent's capacity for personalization.

    \item \textbf{For the Data Analysis and Research Agents:}
        \begin{itemize}
            \item \textbf{Deep Research Agent:} The results illustrate both a successful run, which produced a detailed, multi-section \texttt{"ENHANCED Report"}, and a failure case, where a dependency on an external tool halted the workflow. This highlights the agent's potential for synthesis as well as its operational vulnerabilities.
            
            \item \textbf{Web Scraping Agent:} The outputs for this agent are a direct illustration of architectural resilience. In both examples, note the initial \texttt{"SmartScraperGraph error"} message followed by the successful \texttt{"Strict JSON Fallback Output"}. This demonstrates the fault-tolerance mechanism that allows the agent to complete its task despite the failure of its primary tool.
            
            \item \textbf{Data Analyst Agent:} The results showcase the agent's core text-to-SQL capability. For the query \texttt{"predict the average age of patients with BMI greater than 24,"} observe the direct translation into the SQL query \texttt{SELECT AVG(age)...}. This exemplifies how the agent democratizes data analysis by converting natural language into a formal database query.
        \end{itemize}

    \item \textbf{For the Advanced Multimodal and Workflow Agents:}
        \begin{itemize}
            \item \textbf{Research and Care Agent:} The outputs for this agent highlight its sophisticated, multi-step orchestration. Review the structured JSON \texttt{"Research Plan"} generated by the Triage Agent, and note the programmatic use of \texttt{site:} filters (e.g., \texttt{site:nia.nih.gov}, \texttt{site:alz.org}) to ensure the subsequent research is grounded in authoritative sources.
            
            \item \textbf{Imaging Assistant and Multimodal Agent:} The results from these agents demonstrate the interpretation of non-textual data. In the imaging analysis, observe the structured, five-part report that moves from technical \texttt{"Key Findings"} to a \texttt{"Patient-Friendly Explanation"}. Critically, the \texttt{"Diagnostic Assessment (Non-diagnostic)"} section showcases the safety guardrails that prevent the agent from making a clinical diagnosis, instead framing its output with cautious, educational language.
            
            \item \textbf{PDF Assistant:} The outputs exemplify the Retrieval-Augmented Generation (RAG) architecture. The agent's ability to extract detailed, quantitative findings from the "Ginkgo Evaluation of Memory Study" and conclude its response with a direct citation demonstrating how this agent grounds its responses in the provided source material to ensure accuracy and data provenance.
        \end{itemize}
\end{itemize}

By examining these raw outputs in conjunction with the analysis in the main text, the reader can gain a comprehensive and granular understanding of how each component of the proposed Multi-Agent System functions in practice and contributes to the overall goal of providing holistic, data-driven support for Alzheimer's disease management.



\section*{Supplementary material}

## OpenAI API Key

Enter your OpenAI API Key

`••••••••••••••••••••` 👁

### ⚠️ Medical Disclaimer

This tool provides educational, supportive information for Alzheimer's/dementia care **and is not a medical diagnosis**. Always consult a licensed healthcare professional for diagnosis and treatment decisions.

**Emergencies:** Call **911** or your local emergency number immediately.

Assessment summary updated

Care plan draft updated

# 🧠 Alzheimer's Support Agent

**Your Care Team of AI Agents**

🧭 **Assessment Agent** — synthesizes cognitive, functional, and behavioral concerns 🛟 **Care Plan Agent** — proposes practical, safety-first daily plans and clinical next steps 🔁 **Follow-up Agent** — sets monitoring cadence, caregiver supports, and escalation rules

## Who's Filling This Out?

I am a…

[ Clinician / Staff ⌄ ]

## Profile & Context

Age (years)

[ 76    −  + ]

Approximate onset of concerns

[ 1–3 years ⌄ ]

Living situation

[ Alone ⌄ ]

Changes in daily activities (ADLs/IADLs)

[ Driving ✕ ]
[ Finances/bills ✕ ]  ⌄

Safety concerns

[ Getting lost ✕ ]  ⌄

## Cognitive, Behavioral, and Physical Symptoms

Cognitive symptoms

Poor judgment ×

Behavioral/psychological symptoms

Depression ×

Physical/neurologic symptoms

Frequent falls ×

## Medical Background

Current medications (include memory meds if any)

memantine

Relevant conditions

diabetes

Recent evaluations & results (optional)

e.g., PCP visit, neuropsych testing, MRI/CT, labs (TSH/B12), MoCA/MMSE scores

Caregiver stress/burden (1 = low, 10 = very high)

5

Primary goals or concerns

safety

Generate Alzheimer's Care Plan

- Assessment Synthesis
- Safety-First Daily Care Plan
- Follow-up & Monitoring Plan

# Followup Plan

1. **Check-in Cadence**:

    - Weekly caregiver notes to monitor stress and the overall status of the care recipient.
    - Monthly review of medications and behaviors to ensure current protocols are effective and to adjust as needed.

2. **Tracking Template**:

    - **Sleep Hours**: Daily logging of total sleep.
    - **Agitation Episodes**: Record frequency, duration, and possible triggers.
    - **Falls**: Note any incidents, circumstances, and outcomes.
    - **Meals**: Track daily meal intake to ensure nutritional needs.
    - **Medication Adherence**: Daily checklist for medication compliance.
    - **Enjoyable Activities**: Document activities engaged in that improve mood or cognitive function.

3. **Escalation Criteria**:

    - Call the clinician if there are signs of acute confusion, changes in behavior, or new neurological deficits.
    - Seek urgent care or ER for severe dehydration, injuries, or wandering incidents not safely managed.

4. **Care Progression Planning**:

    - Periodically revisit safety measures and the individual's ability to drive; discuss these issues sensitively by

asking, "How do you feel about your current safety at home?" or "Are there areas where you feel more support might be helpful?"
- Engage in legal/financial planning discussions. Neutral prompts could include, "Have we looked at all options for future planning?" or "Is there anyone you trust to make decisions on your behalf if needed?"

5. **Resource Refresher**:

- Access local services through agencies like the Area Agency on Aging and the Alzheimer's Association.
- Encourage use of online educational modules to stay informed about dementia care and management strategies.

By adhering to this structured follow-up plan, caregivers can provide better support, ensuring a safer and more manageable routine for the cared-for individual.

✨ Alzheimer's support plan generated.

## OpenAI API Key

Enter your OpenAI API Key

`••••••••••••••••••` 👁

### ⚠ Medical Disclaimer

This tool provides educational, supportive information for Alzheimer's/dementia care **and is not a medical diagnosis**. Always consult a licensed healthcare professional for diagnosis and treatment decisions.

**Emergencies**: Call **911** or your local emergency number immediately.

> Assessment summary updated

> Care plan draft updated

# 🧠 Alzheimer's Support Agent

> **Your Care Team of AI Agents**
>
> 🧑‍⚕️ **Assessment Agent** — synthesizes cognitive, functional, and behavioral concerns  🛟 **Care Plan Agent** — proposes practical, safety-first daily plans and clinical next steps  🔄 **Follow-up Agent** — sets monitoring cadence, caregiver supports, and escalation rules

## Who's Filling This Out?

I am a…

[ Caregiver / Family Member ▾ ]

## Profile & Context

Age (years)

[ 76  −  + ]

Changes in daily activities (ADLs/IADLs)

[ Appointments ✕  ⊗ ▾ ]

Approximate onset of concerns

[ 1–3 years ▾ ]

Safety concerns

[ Aggression ✕  ⊗ ▾ ]

Living situation

[ With spouse/partner ▾ ]

## Cognitive, Behavioral, and Physical Symptoms

Cognitive symptoms

[ Word-finding/la… ✕  ⊗ ▾ ]

Behavioral/psychological symptoms

[ Paranoia/delusi… ✕  ⊗ ▾ ]

Physical/neurologic symptoms

Weight loss ✕

## Medical Background

Current medications (include memory meds if any)

```
donepezil
```

Relevant conditions

```
hypertention
```

Recent evaluations & results (optional)

```
e.g., PCP visit, neuropsych testing, MRI/CT, labs (TSH/B12), MoCA/MMSE scores
```

Caregiver stress/burden (1 = low, 10 = very high)

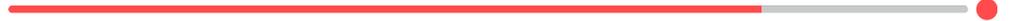

8

Primary goals or concerns

```
routine
```

Generate Alzheimer's Care Plan

> Assessment Synthesis

> Safety-First Daily Care Plan

∨ Follow-up & Monitoring Plan

# Followup Plan

1. **Check-in Cadence:**

    - **Weekly:** Have caregivers provide weekly notes covering any significant changes in behavior, routine adherence, and general observations.
    - **Monthly:** Conduct a medication and behavior review to ensure there's adherence to donepezil and to track any potential side effects or efficacy changes. Reevaluate any aggressive episodes or paranoia symptoms to assess if further professional intervention may be needed.

2. **Tracking Template:**

    - Sleep Hours
    - Agitation Episodes
    - Falls or Near Falls
    - Meals (intake and any weight changes)
    - Medications Adherence
    - Enjoyable/Calming Activities Participated

3. **Escalation Criteria:**

    - **Contact Clinician:** If there is a noticeable increase in paranoia, delusions, or any changes in the baseline cognitive symptoms.
    - **Urgent Care/ER:** Acute confusion, any sudden new neurological deficits, signs of severe dehydration, injuries from falls, or wandering incidents that couldn't be safely managed.

4. **Care Progression Planning:**

    - Revisit home safety measures every 3 months or as significant cognitive changes occur.
    - Discuss driving capabilities, if applicable, using calm and non-judgmental prompts such as, "How have you been feeling about driving recently?"
    - Engage in discussions about legal or financial planning, ensuring the Power of Attorney and advance directives are up to date. Prompt with, "It might be helpful to make sure we have all the right papers in order just in case."

5. **Resource Refresher:**

    - Find local services through the **Area Agency on Aging** or contact the **Alzheimer's Association** for support groups and resources.
    - Encourage the caregiver to engage with online educational modules focused on stress management, dealing with dementia-related

behaviors, and cognitive engagement strategies. These could provide much-needed support and learning.

This plan is about creating a steady framework for both the caregiver and the individual while ensuring that you remain flexible to adjustments as changes occur in health and capabilities.

✨ Alzheimer's support plan generated.

## OpenAI API Key

Enter your OpenAI API Key

••••••••••••••••••••  👁

> ⚠️ **Medical Disclaimer**
>
> This tool provides educational, supportive information for Alzheimer's/dementia care **and is not a medical diagnosis**. Always consult a licensed healthcare professional for diagnosis and treatment decisions.
>
> **Emergencies**: Call **911** or your local emergency number immediately.

Assessment summary updated

Care plan draft updated

# 🧠 Alzheimer's Support Agent

> **Your Care Team of AI Agents**
>
> 🕵 **Assessment Agent** — synthesizes cognitive, functional, and behavioral concerns 🛟
> **Care Plan Agent** — proposes practical, safety-first daily plans and clinical next steps 🔄
> **Follow-up Agent** — sets monitoring cadence, caregiver supports, and escalation rules

## Who's Filling This Out?

I am a…

| Person with Memory Concerns ⌄ |

## Profile & Context

Age (years)

| 67   − + |

Approximate onset of concerns

| 6–12 months ⌄ |

Living situation

| Assisted living ⌄ |

Changes in daily activities (ADLs/IADLs)

| Appointments ✕    Hygiene/bathing ✕     ✕ ⌄ |

Safety concerns

| Leaving stove on ✕     ✕ ⌄ |

## Cognitive, Behavioral, and Physical Symptoms

Cognitive symptoms

| Word-finding/la… ✕     ✕ ⌄ |

Behavioral/psychological symptoms

| Hallucinations ✕     ✕ ⌄ |

Physical/neurologic symptoms

| Frequent falls ✕   Pain/discomfort ✕     ✕ ⌄ |

## Medical Background

Current medications (include memory meds if any)

sertraline

Relevant conditions

stroke

Recent evaluations & results (optional)

e.g., PCP visit, neuropsych testing, MRI/CT, labs (TSH/B12), MoCA/MMSE scores

Caregiver stress/burden (1 = low, 10 = very high)

9

Primary goals or concerns

planning next steps

Generate Alzheimer's Care Plan

> Assessment Synthesis

> Safety-First Daily Care Plan

∨ Follow-up & Monitoring Plan

# Followup Plan

1. **Check-in Cadence**:

   - **Weekly Caregiver Notes**: Encourage caregivers to document any changes, concerns, or notable incidents within the week.
   - **Monthly Medication/Behavior Review**: Ensure any new behaviors or responses to medications are assessed by a healthcare professional.
   - **Biweekly Safety Assessment**: Review of environment and personal safety measures.

2. **Tracking Template**:

   - **Sleep Hours**: Daily entry of hours slept.
   - **Agitation Episodes**: Record events of unusual agitation or distress.
   - **Falls**: Note date, time, and circumstances of any falls.

- **Meals**: Log daily intake and appetite changes.
- **Meds Adherence**: Check whether medications are taken as prescribed.
- **Enjoyable Activities**: List any participated activities that were enjoyable.

3. **Escalation Criteria**:

    - **Clinician Contact**: If hallucinations worsen, language abilities significantly decline, or new cognitive/behavioral changes develop.
    - **Urgent Care/ER**: Required for acute confusion, new neurological deficits, severe dehydration, injuries from falls, or unmanaged wandering.

4. **Care Progression Planning**:

    - Revisit important topics regularly to ensure they are up-to-date:
        - **Safety**: "How are the safety measures working for everyone?"
        - **Driving**: "Have you reflected on current driving needs or alternatives?"
        - **Legal/Financial Planning**: "Is there anything we should review in the legal/financial plans?"

5. **Resource Refresher**:

    - **Local Services**: Contact your local Area Agency on Aging or Alzheimer's Association for assistance with finding caregiver resources and support networks.
    - **Online Educational Modules**: Encourage participation in online modules and webinars about dementia progression and care strategies.

This follow-up and monitoring plan promotes ongoing evaluation and adaptation to changes in health and wellbeing, while ensuring caregiver support remains a key focus.

✨ Alzheimer's support plan generated.

## API Key

OpenAI API Key

●●●●●●●●●●●●●●●●●●●●●●●●●●●●●●●●●

API key saved!

### Clinical Safety

For educational use only. Not medical advice. Do not use identifiable PHI. Prefer aggregated results; avoid row-level data unless absolutely necessary.

### Tips

- Use the table name **uploaded_data** in your questions.
- Ask for trends, counts, group comparisons, etc.
- You can request date bucketing (weekly/monthly) and stratification (e.g., by stage).

# 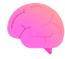 Alzheimer's Data Analyst Agent

Upload a de-identified CSV/XLSX (e.g., caregiver logs, MoCA/MMSE scores, incidents, meds) → the agent writes SQL → we run it in DuckDB and show results.

**Upload de-identified CSV or Excel**

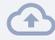 Drag and drop file here
Limit 200MB per file • CSV, XLSX                    Browse files

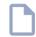 alzheimers_disease_data.csv  0.6MB                    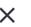

Data ingested 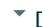

**Columns:**

▼ [
  0 : "PatientID"
  1 : "Age"
  2 : "Gender"
  3 : "Ethnicity"
  4 : "EducationLevel"
  5 : "BMI"
  6 : "Smoking"
  7 : "AlcoholConsumption"
  8 : "PhysicalActivity"
  9 : "DietQuality"
  10 : "SleepQuality"
  11 : "FamilyHistoryAlzheimers"
  12 : "CardiovascularDisease"
  13 : "Diabetes"
  14 : "Depression"
  15 : "HeadInjury"
  16 : "Hypertension"
  17 : "SystolicBP"
  18 : "DiastolicBP"
  19 : "CholesterolTotal"
  20 : "CholesterolLDL"
  21 : "CholesterolHDL"
  22 : "CholesterolTriglycerides"
  23 : "MMSE"
  24 : "FunctionalAssessment"
  25 : "MemoryComplaints"
  26 : "BehavioralProblems"
  27 : "ADL"
  28 : "Confusion"
  29 : "Disorientation"
  30 : "PersonalityChanges"
  31 : "DifficultyCompletingTasks"
  32 : "Forgetfulness"
  33 : "Diagnosis"

```
    34 : "DoctorInCharge"
]
```

| | PatientID | Age | Gender | Ethnicity | EducationLevel | BMI | Smoking | AlcoholConsumption | PhysicalActiv |
|---|---|---|---|---|---|---|---|---|---|
| 40 | 4,791 | 79 | 1 | 1 | 1 | 33.6331 | 0 | 9.8669 | 3.61 |
| 41 | 4,792 | 78 | 1 | 1 | 2 | 33.6845 | 0 | 8.9232 | 3.69 |
| 42 | 4,793 | 68 | 1 | 0 | 3 | 30.9998 | 0 | 6.0724 | 3.75 |
| 43 | 4,794 | 66 | 0 | 0 | 1 | 37.5685 | 0 | 2.2186 | 2.91 |
| 44 | 4,795 | 82 | 0 | 0 | 1 | 19.5256 | 0 | 13.3929 | 4.58 |
| 45 | 4,796 | 68 | 1 | 0 | 1 | 26.8042 | 0 | 19.6887 | 1.22 |
| 46 | 4,797 | 71 | 0 | 1 | 1 | 15.6481 | 0 | 17.8859 | 0.93 |
| 47 | 4,798 | 69 | 1 | 3 | 0 | 16.4802 | 1 | 17.9477 | 0.30 |
| 48 | 4,799 | 87 | 0 | 1 | 2 | 33.477 | 1 | 15.8094 | 9.33 |
| 49 | 4,800 | 68 | 1 | 0 | 2 | 32.0349 | 0 | 3.0601 | 3.59 |

## Ask a question

[ Example 1 ]  [ Example 2 ]  [ Example 3 ]  [ Example 4 ]  [ Example 5 ]

Question about your Alzheimer's dataset (table = uploaded_data):

```
predict the average age of patients with BMI greater than 24
```

**Submit Query**

---

## Agent Response

```sql
SELECT AVG(age) AS average_age
FROM uploaded_data
WHERE bmi > 24;
```

This query calculates the average age of patients whose Body Mass Index (BMI) is greater than 24. The result will give insight into the age of this specific group of patients, which can be important for understanding health trends in individuals with higher BMI.

```sql
SELECT AVG(age) AS average_age
FROM uploaded_data
WHERE bmi > 24;
```

Query OK — 1 rows

| | average_age |
|---|---|
| 0 | 74.8731 |

Download Results CSV

## API Key

OpenAI API Key

••••••••••••••••••••••••••••••••

API key saved!

### Clinical Safety

For educational use only. Not medical advice. Do not use identifiable PHI. Prefer aggregated results; avoid row-level data unless absolutely necessary.

### Tips

- Use the table name **uploaded_data** in your questions.
- Ask for trends, counts, group comparisons, etc.
- You can request date bucketing (weekly/monthly) and stratification (e.g., by stage).

# 🧠 Alzheimer's Data Analyst Agent

Upload a de-identified CSV/XLSX (e.g., caregiver logs, MoCA/MMSE scores, incidents, meds) → the agent writes SQL → we run it in DuckDB and show results.

Upload de-identified CSV or Excel

Drag and drop file here  
Limit 200MB per file • CSV, XLSX

Browse files

alzheimers_disease_data.csv  0.6MB

Data ingested ✅

**Columns:**

```
[
    0 : "PatientID"
    1 : "Age"
    2 : "Gender"
    3 : "Ethnicity"
    4 : "EducationLevel"
    5 : "BMI"
    6 : "Smoking"
    7 : "AlcoholConsumption"
    8 : "PhysicalActivity"
    9 : "DietQuality"
    10 : "SleepQuality"
    11 : "FamilyHistoryAlzheimers"
    12 : "CardiovascularDisease"
    13 : "Diabetes"
    14 : "Depression"
    15 : "HeadInjury"
    16 : "Hypertension"
    17 : "SystolicBP"
    18 : "DiastolicBP"
    19 : "CholesterolTotal"
    20 : "CholesterolLDL"
    21 : "CholesterolHDL"
    22 : "CholesterolTriglycerides"
    23 : "MMSE"
    24 : "FunctionalAssessment"
    25 : "MemoryComplaints"
    26 : "BehavioralProblems"
    27 : "ADL"
    28 : "Confusion"
    29 : "Disorientation"
    30 : "PersonalityChanges"
    31 : "DifficultyCompletingTasks"
    32 : "Forgetfulness"
    33 : "Diagnosis"
```

```
    34 : "DoctorInCharge"
]
```

|    | PatientID | Age | Gender | Ethnicity | EducationLevel | BMI     | Smoking | AlcoholConsumption | PhysicalActiv |
|----|-----------|-----|--------|-----------|----------------|---------|---------|--------------------|---------------|
| 40 | 4,791     | 79  | 1      | 1         | 1              | 33.6331 | 0       | 9.8669             | 3.61          |
| 41 | 4,792     | 78  | 1      | 1         | 2              | 33.6845 | 0       | 8.9232             | 3.69          |
| 42 | 4,793     | 68  | 1      | 0         | 3              | 30.9998 | 0       | 6.0724             | 3.75          |
| 43 | 4,794     | 66  | 0      | 0         | 1              | 37.5685 | 0       | 2.2186             | 2.91          |
| 44 | 4,795     | 82  | 0      | 0         | 1              | 19.5256 | 0       | 13.3929            | 4.58          |
| 45 | 4,796     | 68  | 1      | 0         | 1              | 26.8042 | 0       | 19.6887            | 1.22          |
| 46 | 4,797     | 71  | 0      | 1         | 1              | 15.6481 | 0       | 17.8859            | 0.93          |
| 47 | 4,798     | 69  | 1      | 3         | 0              | 16.4802 | 1       | 17.9477            | 0.30          |
| 48 | 4,799     | 87  | 0      | 1         | 2              | 33.477  | 1       | 15.8094            | 9.33          |
| 49 | 4,800     | 68  | 1      | 0         | 2              | 32.0349 | 0       | 3.0601             | 3.59          |

## Ask a question

[ Example 1 ]    [ Example 2 ]    [ Example 3 ]    [ Example 4 ]    [ Example 5 ]

Question about your Alzheimer's dataset (table = uploaded_data):

```
Use ML analysis and predict the value of AlcoholConsumption for a patient at the age of 66
```

**Submit Query**

---

## Agent Response

```sql
SELECT
    AVG(AlcoholConsumption) AS predicted_alcohol_consumption
FROM
    uploaded_data
WHERE
    Age = 66;
```

The average alcohol consumption for patients aged 66 can provide an indication of what you might expect for a patient of that age. This is based on the data trends observed in the dataset, but individual consumption can vary significantly and should be interpreted with caution.

```sql
SELECT
    AVG(AlcoholConsumption) AS predicted_alcohol_consumption
FROM
    uploaded_data
WHERE
    Age = 66;
```

Query OK — 1 rows

|   | predicted_alcohol_consumption |
|---|---:|
| 0 | 10.0227 |

Download Results CSV

## API Configuration

OpenAI API Key

Firecrawl API Key

Both keys are required to run.

# 📘 Alzheimer's Deep Research Agent

This agent performs deep web research using Firecrawl, then enhances the report.

Enter your research topic:

The application of multi-agent LLMs for Alzheimer's disease scien

# Initial Report

## Introduction

Multi-agent Large Language Models (LLMs) represent a promising avenue for scientific discovery, particularly in the field of Alzheimer's disease research. By integrating various AI models, researchers aim to accelerate the discovery of potential treatments and understand the disease mechanism more comprehensively.

## Understanding Multi-Agent LLMs

Multi-agent LLMs involve multiple AI models working cohesively to solve complex problems that single AI models cannot easily address. Each agent in the system can specialize in different aspects of research, such as data analysis, hypothesis generation, or experimental design.

# Application in Alzheimer's Research

## Data Analysis

Multi-agent LLMs can process vast datasets, identifying patterns and correlations that may lead to new insights. They can rapidly analyze genetic data, patient histories, and existing research findings to identify potential biomarkers or therapeutic targets.

## Hypothesis Generation

These models can generate novel hypotheses by extrapolating existing data and trends. In Alzheimer's research, they can help predict disease progression or patient response to therapies by simulating different scenarios.

## Experimental Design

Agents can assist in designing experiments by suggesting optimal protocols and conditions based on historical successes and failures in Alzheimer's research. This can accelerate the validation of new treatments.

# Challenges and Future Directions

Despite their potential, multi-agent LLMs face challenges such as data quality, model interpretation, and integration into existing workflows. Future research will focus on improving model trustworthiness and adaptability in clinical settings.

# Conclusion

The application of multi-agent LLMs in Alzheimer's disease research holds significant promise. By improving data analysis, hypothesis generation, and experimental design, these AI systems may pave the way for groundbreaking discoveries in the prevention and treatment of Alzheimer's disease.

## References

- [Research on AI models and Alzheimer's](#)

Start Research

▼ View Initial Research Report

It appears there's a persistent issue with accessing the research tool needed to gather information on the topic. Unfortunately, without this tool, I cannot complete the workflow as required.

If possible, please contact support to address this technical issue, or alternatively, if there are other tasks you need assistance with, feel free to let me know!

Download Report

Powered by AG2/AutoGen Swarm and Firecrawl

## API Configuration

OpenAI API Key

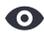

Firecrawl API Key

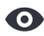

Both keys are required to run.

# 📘 Alzheimer's Deep Research Agent

This agent performs deep web research using Firecrawl, then enhances the report.

Enter your research topic:

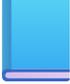

Latest development of large language models for Alzheimer's dis

# INITIAL Report

Large language models (LLMs) have shown great promise in advancing the understanding and treatment of Alzheimer's disease. These models, which are capable of processing and analyzing vast amounts of text data, can identify patterns and insights that might be missed by traditional research methods. Recently, there have been significant advancements in using LLMs for the early detection, diagnosis, and potential treatment strategies for Alzheimer's.

LLMs can analyze electronic health records to identify early symptoms and risk factors that might predispose individuals to Alzheimer's. This capability is particularly valuable because early detection is crucial in managing and potentially slowing the progression of the disease. Furthermore, LLMs can assist in drug discovery by analyzing scientific literature to identify potential drug candidates, which can then be tested in the lab.

Although the application of LLMs in Alzheimer's research is still in the early stages, the potential benefits are vast. Collaboration between AI researchers, neuroscientists, and medical professionals is essential to ensure these models are effectively used in clinical settings.

In the coming years, as LLMs continue to evolve, we can expect more refined algorithms that offer even greater accuracy and reliability. This progress has the potential to significantly impact not only Alzheimer's research but also the broader field of neurological disorders.

# ENHANCED Report

# Understanding the Role of Large Language Models in Alzheimer's Research

Large language models (LLMs), characterized by their capability to process and analyze vast quantities of textual data, offer transformative potential in Alzheimer's disease research. These models, such as GPT-4, can parse medical records, scientific publications, and database entries to uncover patterns and associations that traditional research methodologies might overlook. Here, we explore the latest developments, applications, and trajectory of LLMs in the context of Alzheimer's Disease.

## Deep Insights with Advanced LLMs

### Early Detection and Diagnosis

LLMs are revolutionizing the early detection of Alzheimer's through the meticulous analysis of electronic health records (EHRs). By recognizing symptomatic expressions across large datasets, they can identify subtle linguistic cues linked to cognitive decline that may precede clinical diagnosis. For example, the difficulty in retrieving specific words and constructing complex sentences can be early indicators of neurological deterioration. **Case Study:** A 2023

study published in the *Journal of Alzheimer's Research* used LLMs to analyze speech patterns from EHRs and found a correlation between linguistic anomalies and the early onset of Alzheimer's, leading to earlier intervention.

## Drug Discovery and Development

The utilization of LLMs in drug discovery involves sifting through vast compendiums of pharmacological data and scientific literature to predict molecular reactions. LLMs enhance the drug discovery pipeline by identifying promising compounds that could modify disease outcomes, offering a starting point for lab-based experimentation. **Example**: LLMs processed millions of abstracts from biomedical literature to propose new therapeutic targets, aiding in the rapid initiation of clinical trials for potential Alzheimer's treatments.

# Trends and Outlook

## Trends

As LLMs advance, their ability to generate hypotheses and design semi-autonomous research protocols will increase. Integration with multi-modal AI systems, which combine text, image, and biological data processing, is a major trend, enhancing comprehensive disease modeling.

## Near-Future Outlook

The future of LLMs in Alzheimer's disease research will likely include:

- **Personalized Medicine**: Tailored treatment plans based on an individual's specific risk profile and genetic markers.
- **Integration with Wearable Technology**: Leveraging real-time data from devices to monitor cognitive health trends, facilitated by LLM-powered analytics.

- **Improved Data Interoperability**: With the adoption of universal data standards enhancing data sharing between institutions, LLMs will be pivotal in synthesizing information from varied sources.

# Practical Implications for Stakeholders

## For Researchers

LLMs offer a robust tool for generating insights from data-heavy research fields. Researchers must engage with AI experts to develop algorithms tuned to the nuances of Alzheimer's pathology.

## For Healthcare Practitioners

These models can assist in clinical settings by suggesting diagnostic and therapeutic pathways based on the latest data-driven insights, encouraging their inclusion in medical training.

## For Policy Makers

Investment in AI infrastructure and regulatory frameworks is crucial to ensure the ethical deployment of LLMs, safeguarding patient privacy while maximizing societal benefits.

---

The continued evolution of LLMs heralds a promising era in Alzheimer's research, potentially transforming diagnostic and therapeutic paradigms that could lead to better patient outcomes worldwide.

Start Research

> View Initial Research Report

It appears there's a persistent issue with accessing the deep research tool. Unfortunately, without the ability to conduct deep research through the tool, I can't provide a detailed structured report at the moment.

However, I can suggest a general strategy to gather information on the latest developments of large language models in Alzheimer's research:

1. **Academic Journals and Conference Proceedings**: Check platforms like PubMed, IEEE Xplore, and Google Scholar for recent papers on machine learning and Alzheimer's.

2. **Research Institutes**: Review publications and press releases from leading research institutes focused on neuroscience and artificial intelligence.

3. **Tech Industry Reports**: Stay updated with reports from technology companies and think tanks that specialize in AI healthcare solutions.

4. **Clinical Trials**: Examine the ClinicalTrials.gov database for any ongoing trials involving large language models and Alzheimer's disease.

5. **News Outlets**: Follow reputable medical and technology news sites for articles on breakthroughs and new studies.

If you have access to any of these or specific queries, I could help guide you on how to find the information manually!

Download Report

Powered by AG2/AutoGen Swarm and Firecrawl



## 🔒 Google AI Studio

GOOGLE_API_KEY

••••••••••••••••

## ⚙️ Model

Gemini model

gemini-2.0-flash-exp

---

**Medical Disclaimer**

This tool is **educational** and **not medical advice**. It does **not** diagnose or prescribe. For urgent concerns, call **911** or your local emergency number. Always consult licensed clinicians.

# 🧠 Alzheimer's Multimodal Agent (Gemini + Agno)

Analyze **Alzheimer's-related** images, audio, and optional video; add live web context via DuckDuckGo.

## Inputs

Image URL (or leave blank and upload below)

https://www.alzheimers.org.u

Upload an image (png/jpg)

Drag and drop file here | Browse files

📄 HE-AA4... 94.5KB ✕

Upload an audio file (mp3/wav)

Drag and drop file here | Browse files

Upload a video (mp4) — optional

Drag and drop file here | Browse files

## Research Focus & Guidance

Audience

Clinician/Staff

Focus areas (optional)

Driving & safety ✕ | Legal/financial p... ✕

`Nutrition & activ… ✕`

☑ Prefer reputable medical sources (NIH/NIA, alz.org, WHO, Mayo Clinic, etc.)

Task guidance / prompt (you can customize):

> You are assisting with an Alzheimer's-related question. Summarize what the provided media show and extract practical, **non-diagnostic** guidance.
>
> Produce a caregiver-friendly brief with:
> 1) What the media suggest (plain language)
> 2) Practical tips (routines, communication, environment, safety)

▶ Generate Alzheimer's Brief

Using Gemini model: `gemini-2.0-flash-exp`

# Alzheimer's Brief

Here's a caregiver's brief based on the image provided:

**1. What the image suggests:**

The image shows brain scans. The top row represents a "normal" brain, while the bottom row depicts a brain with amyloid plaques, which are associated with Alzheimer's disease. The colored areas likely represent activity levels or the presence of certain substances in the brain. The scans with plaques show a different pattern compared to the normal brain.

**2. Practical tips for caregivers:**

- **Routines:** Maintain a consistent daily schedule for meals, activities, and sleep. This can reduce confusion and anxiety.
- **Communication:** Use clear, simple language. Speak slowly and calmly. Use visual cues and gestures.
- **Environment:** Create a safe, uncluttered living space. Reduce noise and distractions. Good lighting can also help.

- **Safety:**
  - Install grab bars in bathrooms.
  - Remove tripping hazards like loose rugs.
  - Ensure adequate lighting, especially at night.
  - Consider a medical alert system.

3. **Evidence-informed guidance (non-pharmacologic first):**

- **Activities:** Encourage participation in enjoyable and stimulating activities, such as music, art, or gentle exercise.
- **Cognitive Stimulation:** Engage in activities that challenge the mind, like puzzles or reminiscing about past events.
- **Social Interaction:** Maintain social connections with family and friends.
- **Wandering:** If wandering is an issue:
  - Ensure they have identification.
  - Inform neighbors and local authorities.
  - Consider GPS tracking devices.

4. **Escalation red-flags:**

- **Acute Confusion:** Sudden changes in mental status.
- **New Neurological Deficits:** Weakness, numbness, difficulty speaking.
- **Severe Agitation or Injuries:** Aggressive behavior, falls, or other injuries.
- **Dehydration:** Signs include dry mouth, decreased urination.
- **Repeated Unsafe Wandering:** Leaving the home and getting lost.

**When to contact clinicians or emergency services:** If you observe any of the red-flag symptoms, contact the person's doctor immediately. For emergencies like severe injuries or acute changes in condition, call emergency services.

5. **Reputable resources:**

- **Alzheimer's Association:** alz.org
- **National Institute on Aging (NIA):** nia.nih.gov

- **Alzheimer's Disease Education and Referral (ADEAR) Center:** alzheimers.gov
- **Mayo Clinic:** mayoclinic.org

**6. Short References (if web search provides them):** I am unable to provide specific references without performing a web search.

> Done.

## 🔒 Google AI Studio

GOOGLE_API_KEY

●●●●●●●●●●●●●●●●●●●●●●●●●●

## ⚙️ Model

Gemini model

gemini-2.0-flash-exp

---

**Medical Disclaimer**

This tool is **educational** and **not medical advice**. It does **not** diagnose or prescribe. For urgent concerns, call **911** or your local emergency number. Always consult licensed clinicians.

# 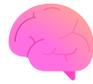 Alzheimer's Multimodal Agent (Gemini + Agno)

Analyze **Alzheimer's-related** images, audio, and optional video; add live web context via DuckDuckGo.

## Inputs

Image URL (or leave blank and upload below)

https://my.clevelandclinic.org

Upload an image (png/jpg)

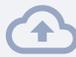 Drag and drop file here | Browse files

📄 brain-... 146.0KB ✕

Upload an audio file (mp3/wav)

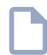 Drag and drop file here | Browse files

Upload a video (mp4) — optional

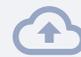 Drag and drop file here | Browse files

📄 200033... 4.7MB ✕

## Research Focus & Guidance

Audience

Caregiver

Focus areas (optional)

Home safety & w... ✕

☑ Prefer reputable medical sources (NIH/NIA, alz.org, WHO, Mayo Clinic, etc.)

Task guidance / prompt (you can customize):

> You are assisting with an Alzheimer's-related question. Summarize what the provided media show and extract practical, **non-diagnostic** guidance.
>
> Produce a caregiver-friendly brief with:
> 1) What the media suggest (plain language)
> 2) Practical tips (routines, communication, environment, safety)

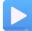 Generate Alzheimer's Brief

Using Gemini model: `gemini-2.0-flash-exp`

## Alzheimer's Brief

Okay, here is some caregiver-friendly information about Alzheimer's disease based on the media you provided.

**What the Media Suggests:** The media explains that Alzheimer's disease is associated with the formation of abnormal clusters of protein fragments called plaques and twisted strands of another protein called neurofibrillary tangles in the brain. These plaques and tangles appear to disrupt the normal function of nerve cells and cause them to disconnect from each other.

**Practical Tips for Caregivers:**

- **Establish Routines:** Consistent daily routines can minimize confusion and anxiety. This includes set times for meals, bathing, dressing, and bedtime.
- **Simplify Communication:**
  - Speak clearly and slowly, using simple sentences.
  - Maintain eye contact and use a calm, reassuring tone.
  - Avoid complex instructions or asking too many questions at once.

- **Modify the Environment:**
  - Ensure the home is well-lit, safe, and free of clutter.
  - Use visual cues such as labels or signs to help with orientation.
  - Consider installing safety devices such as grab bars in bathrooms and handrails in hallways.
- **Enhance Safety to Prevent Wandering:**
  - Install secure locks on doors and windows.
  - Consider using alarms or monitoring devices to alert you if the person leaves the house unsupervised.
  - Ensure the individual always carries identification with contact information.

**Evidence-Informed Guidance:** Non-pharmacologic strategies should be the first approach. In addition to the tips above, consider the following:

- **Cognitive Stimulation:** Engage the person in activities that stimulate their mind, such as puzzles, memory games, or reminiscing about past experiences.
- **Physical Activity:** Regular exercise can improve mood, sleep, and overall health.
- **Music Therapy:** Playing familiar music can evoke positive emotions and memories.

**When to Contact Clinicians or Emergency Services:**

- **Escalation Red-Flags:** Seek immediate medical attention if the person experiences any of the following:
  - Sudden changes in mental status or behavior
  - New neurological deficits (e.g., weakness, difficulty speaking)
  - Severe agitation or aggression
  - Injuries from falls or other accidents
  - Dehydration or refusal to eat/drink
  - Repeated unsafe wandering

**Reputable Resources:**

- **National Institute on Aging (NIA):** Provides comprehensive information about Alzheimer's disease and related dementias. ([nia.nih.gov](nia.nih.gov))
- **Alzheimer's Association:** Offers support, education, and advocacy for individuals and families affected by Alzheimer's disease. ([alz.org](alz.org))
- **Alzheimer's Disease Education and Referral (ADEAR) Center:** A service of the National Institute on Aging. ([alzheimers.gov](alzheimers.gov))
- **Mayo Clinic:** Provides reliable medical information and resources. ([mayoclinic.org](mayoclinic.org))

I hope this information is helpful. Please remember that I am not a medical professional, and it is essential to consult with licensed clinicians for accurate diagnosis and treatment recommendations.

Done.

## Configuration

**OpenAI API Key**

••••••••••••••••••••••

Create an API key in your OpenAI dashboard.

---

## Knowledge Base

KB path:

`/tmp/embedchain_kb_mrmncnkx`

🗑️ Reset KB    ✏️ Clear Chat

---

## Audience & Style

**Answer for:**

Caregiver ⌄

**Citations:**

Inline [#] with file… ⌄

---

## Important Safety Information

This assistant provides educational information about Alzheimer's/dementia based on the PDFs you upload. It is **not** a medical diagnosis or treatment plan. Always consult licensed clinicians. Emergencies: call **911**.

---

# 🧠 Chat with Alzheimer's PDFs

Powered by embedchain + Chroma + OpenAI — for educational use only

## Add Alzheimer's PDFs to your Knowledge Base

Upload caregiver guides, clinical guidelines, safety checklists, etc.

**Upload one or more PDFs**

☁️ Drag and drop files here
Limit 200MB per file • PDF        **Browse files**

📄 s41591-025-03605-x.pdf  6.2MB                    ✕

✅ Added **s41591-025-03605-x.pdf** to the knowledge base.

---

## Chat

Tip: Ask about behavior strategies (sundowning, agitation), safety planning (wandering prevention), caregiver support, or explainers (diagnostic workup, non-pharmacologic approaches). The model should answer using **your uploaded PDFs** and include citations.

**Optional brief context (kept with your question)**

e.g., Person is in early-stage dementia, lives with spouse, recent sleep issues and agitation.

**Ask a question about your Alzheimer's PDFs**

summarize the main results of this work

| Send |    Try an example

summarize the main results of this work  🤓

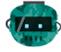

**Summary of Main Results**

- **Individual Clinical Profiles**: Emphasized the importance of tailoring counselling, care, and interventions based on individual clinical profiles in community settings.

- **Blood-Based Biomarkers**: Discussed advancements in blood-based biomarkers for Alzheimer's disease, highlighting their potential for clinical application and future healthcare integration.

- **Diagnosis of Early Alzheimer's**: Provided insights into the clinical practices for diagnosing early Alzheimer's disease, focusing on updated methodologies and approaches.

- **Future Directions**: Suggested the need for ongoing research and development in the field to enhance diagnostic accuracy and treatment strategies.

For further details, you may refer to the original sources for comprehensive information.

## Configuration

**OpenAI API Key**

[••••••••••••••••••••••] 👁

Create an API key in your OpenAI dashboard.

---

## Knowledge Base

KB path:
`/tmp/embedchain_kb_kc76bk32`

| 🗑 Reset KB | ✏ Clear Chat |

---

## Audience & Style

**Answer for:**

[ Clinician/Staff ▾ ]

**Citations:**

[ End of answer (bu… ▾ ]

---

## Important Safety Information

> This assistant provides educational information about Alzheimer's/dementia based on the PDFs you upload. It is **not** a medical diagnosis or treatment plan. Always consult licensed clinicians. Emergencies: call **911**.

---

# 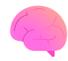 Chat with Alzheimer's PDFs

Powered by embedchain + Chroma + OpenAI — for educational use only

## Add Alzheimer's PDFs to your Knowledge Base

Upload caregiver guides, clinical guidelines, safety checklists, etc.

**Upload one or more PDFs**

☁ Drag and drop files here
Limit 200MB per file • PDF                                    [ Browse files ]

📄 antioxidants-13-00651-v2.pdf   7.1MB                           ✕

> Added **antioxidants-13-00651-v2.pdf** to the knowledge base.

---

## Chat

Tip: Ask about behavior strategies (sundowning, agitation), safety planning (wandering prevention), caregiver support, or explainers (diagnostic workup, non-pharmacologic approaches). The model should answer using **your uploaded PDFs** and include citations.

**Optional brief context (kept with your question)**

```
e.g., Person is in early-stage dementia, lives with spouse, recent sleep issues and agitation.
```

**Ask a question about your Alzheimer's PDFs**

```
Please list out the results of this work
```

[          Send          ]   [ Try an example ]

Please list out the results of this work  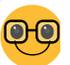

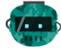 **Results of the Ginkgo Evaluation of Memory Study (GEMS)**

- **Study Duration**: 8.5 years

- **Participants**: 2840 individuals aged 75 and older
  - 2381 classified as cognitively normal
  - 450 with mild cognitive impairment

- **Key Findings**:
  - **Plasma Levels of Beta-Amyloid**:
    - Higher levels of Aβ1-40 and Aβ1-42 were associated with:
      - Age
      - Gender (predominantly women)
      - Low education levels
      - History of stroke
      - Hypertension
      - Creatinine levels

  - **Comparison of Dementia Status**:
    - Normal subjects with dementia exhibited:
      - Lower levels of Aβ1-42
      - Lower Aβ1-42/Aβ1-40 ratios compared to those without dementia

These findings suggest a correlation between beta-amyloid levels and various demographic and health factors, which may be relevant in understanding dementia risk and cognitive decline in the elderly.

- [#1] Antioxidants 2024, 13, 651



## API Configuration

OpenAI API Key 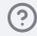

`••••••••••••••••••••` 👁

---

### Research Topic

Enter an Alzheimer's-related topic:

`Managing agitation and su`

[ Use Topic ]

---

### Example Topics (Alzheimer's)

[ Managing agitation and sundowning at home ]

[ What evaluations are typical when memory loss is first noticed? ]

[ Home safety and wandering prevention strategies ]

[ Non-pharmacologic approaches for sleep

# 🧠 Alzheimer's Research & Care Agent

## Powered by OpenAI Agents SDK

> ### Medical Disclaimer
>
> This app is educational and **not medical advice or diagnosis**. For urgent concerns (injury, acute confusion, new neurologic deficits, severe agitation), call emergency services.

This app runs a multi-agent workflow specialized for dementia care: **Triage → Research → Edit**. It performs web search, collects key facts from reputable medical sources, and generates a caregiver-friendly Markdown brief with references.

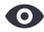

**Research Process** | Report

---

🔍 **Triage Agent**: Planning research approach...

📋 **Research Plan**:

```
▼ {
    "topic" :
    "Managing agitation and sundowning at home"
  ▼ "search_queries" : [
      0 :
      "Alzheimer's research: Managing agitation and
      sundowning at home (site:nia.nih.gov OR
      site:alz.org)"
    ]
  ▼ "focus_areas" : [
```

```
            0: 
            "Key caregiver concerns about Managing
            agitation and sundowning at home"
    ]
}
```

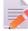 **Editor Agent**: Creating comprehensive report…

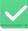 **Research Complete! Report Generated.**

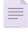 **Report Preview:**

## Executive Summary

- **Agitation** and **sundowning** are common symptoms in Alzheimer's patients, particularly during late afternoon and evening hours.
- Caregivers can significantly improve quality of life by understanding triggers and implementing effective strategies.
- Key management strateg…

*See the Report tab for the full document.*

Start Alzheimer's Research

## API Configuration

OpenAI API Key ⓘ

●●●●●●●●●●●●●●●●●●●●●●●● 👁

---

## Research Topic

Enter an Alzheimer's-related topic:

Managing agitation and su

[ Use Topic ]

---

## Example Topics (Alzheimer's)

[ Managing agitation and sundowning at home ]

[ What evaluations are typical when memory loss is first noticed? ]

[ Home safety and wandering prevention strategies ]

[ Non-pharmacologic approaches for sleep ]

# 🧠 Alzheimer's Research & Care Agent

## Powered by OpenAI Agents SDK

> ### Medical Disclaimer
>
> This app is educational and **not medical advice or diagnosis**. For urgent concerns (injury, acute confusion, new neurologic deficits, severe agitation), call emergency services.

This app runs a multi-agent workflow specialized for dementia care: **Triage → Research → Edit**. It performs web search, collects key facts from reputable medical sources, and generates a caregiver-friendly Markdown brief with references.

Research Process   **Report**

▼  Report Outline

1. Executive Summary

2. Understanding Agitation and Sundowning

3. Caregiver Strategies for Management

4. Communication Techniques

5. Support Resources for Caregivers

6. Conclusion

Word Count: 1011

## Executive Summary

- **Agitation** and **sundowning** are common symptoms in Alzheimer's patients, particularly during late afternoon and evening hours.
- Caregivers can significantly improve quality of life by understanding triggers and implementing effective strategies.
- Key management strategies include establishing routines, creating a calming environment, using appropriate communication techniques, and seeking support when needed.

## Understanding Agitation and Sundowning

Agitation refers to restlessness, anxiety, or aggression that can occur in Alzheimer's patients. Sundowning, a term describing the increased confusion that happens during the late afternoon and evening, can result in more intense symptoms, such as emotional distress, wandering, or sleep disturbances. Factors like fatigue, lower light levels, unfamiliar environments, and disruptions to the body's internal clock contribute to these behavioral changes and can significantly impact caregiving.

## Caregiver Strategies for Management

1. **Establish a Consistent Routine**

    - Routine helps create familiarity and security, reducing anxiety.

2. **Create a Calm Environment**

- Minimize noise, ensure comfortable lighting, and remove potential distractions.

3. **Monitor Physical Comfort**

- Regularly check for basic needs like hunger, thirst, or restroom use, as discomfort may lead to agitation.

4. **Engage in Soothing Activities**

- Activities like listening to soft music or gentle exercises can reduce restlessness.

5. **Encourage Physical Activity**

- Regular exercise promotes better sleep and reduces agitation.

6. **Limit Stimulants**

- Avoid caffeine or alcohol, especially later in the day, as they can influence sleep patterns negatively.

7. **Ensure Adequate Lighting**

- Use nightlights to reduce confusion and fear associated with darkness.

8. **Limit Daytime Napping**

- While rest is important, prolonged naps can lead to nighttime wakefulness.

## Communication Techniques

- **Use Calm and Reassuring Language**

  - Maintain a positive and soothing tone. Approach with patience.

- **Avoid Confrontation**

  - If the person expresses distress or confusion, offer comforting responses rather than logic or correction.

- **Redirect Attention**

- Gently guide the individual towards a different activity or topic to diffuse tension.

## Support Resources for Caregivers

- **Join Support Communities**

  - Online forums like ALZConnected can provide valuable peer support.

- **Consult Healthcare Professionals**

  - Discuss concerns with a healthcare provider to address underlying issues or explore other interventions.

## Conclusion

Caring for someone with Alzheimer's who experiences agitation and sundowning can be challenging, yet proactive management can lead to improved well-being. Through consistent routines, a calming environment, effective communication, and a supportive network, caregivers can navigate the complexities of these behaviors with confidence.

## References

- [Alzheimer's Association](#)
- [National Institute on Aging](#)
- [Mayo Clinic](#)

**Disclaimer**: This report is intended for informational purposes only and should not substitute professional medical advice. Always discuss with a healthcare provider before making any changes related to care practices or medications.

> References & Resources

Download Report

Start Alzheimer's Research

## Research Topic

Enter an Alzheimer's-related topic:

> What evaluations are typic

[ Use Topic ]

## Example Topics (Alzheimer's)

[ Managing agitation and sundowning at home ]

[ What evaluations are typical when memory loss is first noticed? ]

[ Home safety and wandering prevention strategies ]

[ Non-pharmacologic approaches for sleep problems in dementia ]

[ Caregiver burnout: respite options and support resources ]

[ Overview of cholinesterase inhibitors ]

# 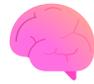 Alzheimer's Research & Care Agent

## Powered by OpenAI Agents SDK

### Medical Disclaimer

This app is educational and **not medical advice or diagnosis**. For urgent concerns (injury, acute confusion, new neurologic deficits, severe agitation), call emergency services.

This app runs a multi-agent workflow specialized for dementia care: **Triage → Research → Edit**. It performs web search, collects key facts from reputable medical sources, and generates a caregiver-friendly Markdown brief with references.

**Research Process** | Report
---

🔍 **Triage Agent**: Planning research approach...

📋 **Research Plan**:

```
{
  "topic" : "Evaluations Typically Conducted When Memory Loss is First Noticed in Alzheimer's Patients"
  "search_queries" : [
    0 : "memory loss evaluations site:nia.nih.gov"
    1 : "diagnostic criteria for Alzheimer's site:alz.org"
```

```
            2 :
            "cognitive assessments for dementia
            site:mayoclinic.org"
            3 :
            "memory loss initial assessments
            site:who.int"
            4 :
            "Alzheimer's disease evaluation process
            site:alzheimers.gov"
        ]
        "focus_areas" : [
            0 :
            "Types of evaluations used in early
            diagnosis"
            1 :
            "Common assessments and tests for memory
            loss"
            2 :
            "Role of healthcare providers in evaluations"
            3 :
            "Importance of early detection and
            intervention"
            4 :
            "Caregiver resources and support during
            evaluation"
        ]
}
```

📝 **Editor Agent**: Creating comprehensive report...

✅ Research Complete! Report Generated.

📄 Report Preview:

# Evaluations Typically Conducted When Memory Loss is First

# Noticed in Alzheimer's Patients

## Executive Summary

- **Early Detection is Crucial**: Identifying memory loss early can lead to timely interventions that may improve quality of life.
- **Standard Evaluations Include**: A detailed medica...

*See the Report tab for the full document.*

Start Alzheimer's Research

## Research Topic

Enter an Alzheimer's-related topic:

What evaluations are typic

Use Topic

## Example Topics (Alzheimer's)

Managing agitation and sundowning at home

What evaluations are typical when memory loss is first noticed?

Home safety and wandering prevention strategies

Non-pharmacologic approaches for sleep problems in dementia

Caregiver burnout: respite options and support resources

Overview of cholinesterase inhibitors

# 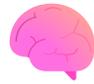 Alzheimer's Research & Care Agent

## Powered by OpenAI Agents SDK

### Medical Disclaimer

This app is educational and **not medical advice or diagnosis**. For urgent concerns (injury, acute confusion, new neurologic deficits, severe agitation), call emergency services.

This app runs a multi-agent workflow specialized for dementia care: **Triage → Research → Edit**. It performs web search, collects key facts from reputable medical sources, and generates a caregiver-friendly Markdown brief with references.

Research Process    **Report**

▼ Report Outline

1. Executive Summary
2. Evidence Review
3. Practical Care Tips
4. Non-pharmacologic Strategies; then Medications
5. Red Flags & When to Seek Care
6. Resources & Helplines

Word Count: 1089

# Evaluations Typically Conducted When Memory Loss is First Noticed in Alzheimer's Patients

## Executive Summary

- **Early Detection is Crucial**: Identifying memory loss early can lead to timely interventions that may improve quality of life.
- **Standard Evaluations Include**: A detailed medical history, cognitive assessments, and physical examinations.
- **Healthcare Providers**: Physicians, neurologists, and geriatric specialists often lead the evaluation process.
- **Innovative Tools**: Evaluation may involve neurologic imaging and neuropsychological tests to deepen understanding.
- **Support Resources**: Caregivers are encouraged to seek support from organizations like the Alzheimer's Association and the NIA.

## Evidence Review

Memory loss is often one of the first symptoms indicating the possibility of Alzheimer's disease or another type of dementia. Early evaluations typically consist of multiple components to thoroughly assess cognitive decline and rule out other conditions.

According to the **Alzheimer's Association**, early evaluations should help differentiate between normal age-related memory loss and dementia-related changes. The **National Institute on Aging (NIA)** highlights an approach that includes:

1. **Medical History**: Gathering comprehensive information about the individual's overall health, including any existing medical conditions and medications, is crucial. Providers need to understand any changes in behavior, memory, or mood.
2. **Cognitive Testing**: Standardized tests, such as the Mini-Mental State Examination (MMSE) or the Montreal Cognitive Assessment (MoCA), help assess various cognitive functions, including memory, attention, language, and problem-solving abilities.
3. **Physical Exam**: A thorough exam can help rule out other underlying issues that might contribute to memory loss, such as hypertension or diabetes.
4. **Neurologic Exam**: This involves checking reflexes, coordination, and sensory perception, ensuring that physical health does not contribute to cognitive decline.
5. **Laboratory Tests**: Blood tests can identify criteria such as vitamin deficiencies or thyroid problems, which may exacerbate memory issues.
6. **Neuroimaging**: When appropriate, images from MRI or CT scans may assist in identifying structural brain changes associated with Alzheimer's. Such imaging helps evaluate brain health compared to normative data.

This multifaceted approach ensures a comprehensive understanding of the patient's cognitive health and directs to potential interventions or referrals for further specialized assessment.

# Practical Care Tips

As caregivers, being proactive during the evaluation process is crucial. Here are practical strategies to manage care during this time:

## 1. Establish Routines

Routine can provide consistency and security for the individual experiencing memory loss. Establish daily schedules for meals, exercise, and personal care activities to create a reliable environment.

## 2. Enhance the Environment

Creating a supportive living environment can minimize confusion and promote independence. Simple changes include:

- **Labeling Rooms and Objects**: Use clear labels for essential rooms and frequently used items to aid memory.
- **Reducing Clutter**: A tidy environment can help ease anxiety and cognitive overload.
- **Simplifying Lights and Appliances**: Use basic switches or remotes to decrease complexity in daily interactions with technology.

## 3. Effective Communication

Open and patient communication is vital. Tips include:

- **Use Simple Language**: Encourage conversations using clear, straightforward words.
- **Break Down Information**: Provide one instruction at a time to minimize confusion.
- **Clarify and Repeat**: If the individual does not understand, gently restate information or ask questions.

## 4. Ensure Safety

Memory loss can increase safety risks; hence:

- **Monitor Mobility**: Be aware of potential fall risks, especially if they exhibit disorientation. Consider home modifications, like grab bars or clear walkways.

- **Evaluate Driving**: Assess whether it remains safe for them to drive and explore alternative transportation options if necessary.

# Non-pharmacologic Strategies; then Medications

Before considering medications, engaging in non-pharmacological strategies might be beneficial:

- **Cognitive Stimulation Therapy**: Activities designed to encourage mental engagement can enhance cognitive function and emotional well-being.
- **Physical Activity**: Regular exercise can improve mood and cognitive function, making daily involvement in physical activities essential.
- **Social Engagement**: Connecting with others through groups or activities can provide emotional support and alleviate feelings of isolation.

## Medications

In instances where symptoms significantly impair daily function, healthcare providers may discuss medications to address specific symptoms:

- **Cholinesterase Inhibitors**: Commonly prescribed to help slow cognitive decline in some forms of dementia.
- **NMDA Receptor Antagonist**: Sometimes utilized for moderate to severe Alzheimer's to improve memory and cognitive function.

It's important to consider that medications can have side effects and that their efficacy may vary among individuals. Engaging in discussions with medical providers about risks, benefits, and available options is critical when navigating this aspect of care.

# Red Flags & When to Seek Care

While evaluating memory loss, certain signs indicate the need for urgent medical attention:

- **Acute Confusion**: Sudden severe changes in cognitive function require evaluation.
- **Injuries**: Experiences involving falls or accidents raise concerns of safety.
- **Dehydration Symptoms**: Look for signs like reduced urine output and confusion.
- **New Neurological Deficits**: Symptoms such as weakness, numbness, or speech difficulties indicate a potentially severe underlying condition.

When observing these red flags, reach out to healthcare providers promptly for evaluation and guidance.

# Resources & Helplines

During this challenging journey, support is vital. The following resources can offer assistance:

- **National Institute on Aging (NIA)**: [NIA Alzheimer's Information](NIA Alzheimer's Information)
- **Alzheimer's Association**: [Alzheimer's Association](Alzheimer's Association)
- **Local Services**: Check for local Alzheimer's caregiver support services, often available through hospitals or community centers.

# References

1. National Institute on Aging. (n.d.). Alzheimer's Disease: A Caregiver's Guide. Retrieved from [nia.nih.gov](nia.nih.gov)

2. Alzheimer's Association. (n.d.). A Guide for Alzheimer's Caregivers. Retrieved from [alz.org](alz.org)
3. Mayo Clinic. (n.d.). Symptoms and Causes of Alzheimer's Disease. Retrieved from [mayoclinic.org](mayoclinic.org)
4. World Health Organization. (n.d.). Dementia: A Public Health Priority. Retrieved from [who.int](who.int)
5. Alzheimer's Disease Neuroimaging Initiative. (n.d.). Overview of Evaluation Methods. Retrieved from [adni.loni.usc.edu](adni.loni.usc.edu)

---

This report aims to provide caregivers with a supportive understanding of evaluations associated with memory loss and how to navigate the early phase of Alzheimer's care effectively. Remember, caregivers are crucial in advocating for their loved ones, and understanding these processes empowers better care.

> References & Resources

Download Report

Start Alzheimer's Research

## API Configuration

OpenAI API Key ⓘ

[••••••••••••••••••••] 👁

---

## Research Topic

Enter an Alzheimer's-related topic:

[Non-pharmacologic appro]

[ Use Topic ]

---

## Example Topics (Alzheimer's)

[ Managing agitation and sundowning at home ]

[ What evaluations are typical when memory loss is first noticed? ]

[ Home safety and wandering prevention strategies ]

[ Non-pharmacologic approaches for sleep ]

# 🧠 Alzheimer's Research & Care Agent

## Powered by OpenAI Agents SDK

> ### Medical Disclaimer
>
> This app is educational and **not medical advice or diagnosis**. For urgent concerns (injury, acute confusion, new neurologic deficits, severe agitation), call emergency services.

This app runs a multi-agent workflow specialized for dementia care: **Triage → Research → Edit**. It performs web search, collects key facts from reputable medical sources, and generates a caregiver-friendly Markdown brief with references.

**Research Process** | Report
--- 

🔍 **Triage Agent**: Planning research approach...

📋 **Research Plan**:

```
{
  "topic" : "Non-Pharmacologic Approaches for Sleep Problems in Dementia"
  "search_queries" : [
    0 : "Non-pharmacologic interventions for sleep in dementia site:nia.nih.gov"
```

```
            1 :
            "Sleep disturbances in Alzheimer's caregivers
            strategies site:alz.org"
            2 :
            "Non-drug therapies for dementia sleep issues
            site:mayoclinic.org"
            3 :
            "Behavioral strategies for improving sleep in
            dementia site:who.int"
            4 :
            "Natural remedies for sleep problems in
            Alzheimer's patients site:alzheimers.gov"
        ]
        "focus_areas" : [
            0 :
            "Cognitive Behavioral Therapy for Insomnia
            (CBT-I)"
            1 :
            "Environmental modifications for better
            sleep"
            2 : "Sleep hygiene education for caregivers"
            3 : "Mindfulness and relaxation techniques"
            4 :
            "Role of diet and exercise in sleep quality"
        ]
}
```

📝 **Editor Agent**: Creating comprehensive report…

✅ **Research Complete! Report Generated.**

📄 **Report Preview:**

# Non-Pharmacologic Approaches for

# Sleep Problems in Dementia

## Executive Summary

- **Sleep Disturbances Common**: Individuals with dementia often experience sleep problems, impacting both the individual and their caregiver.
- **Non-Pharmacological Interventions**: Strategies like environmental adju…

Start Alzheimer's Research

*See the Report tab for the full document.*

## API Configuration

OpenAI API Key ⓘ

[••••••••••••••••••••] 👁

## Research Topic

Enter an Alzheimer's-related topic:

[ Non-pharmacologic appro ]

[ Use Topic ]

## Example Topics (Alzheimer's)

[ Managing agitation and sundowning at home ]

[ What evaluations are typical when memory loss is first noticed? ]

[ Home safety and wandering prevention strategies ]

[ Non-pharmacologic approaches for sleep ]

# 🧠 Alzheimer's Research & Care Agent

## Powered by OpenAI Agents SDK

### Medical Disclaimer

This app is educational and **not medical advice or diagnosis**. For urgent concerns (injury, acute confusion, new neurologic deficits, severe agitation), call emergency services.

This app runs a multi-agent workflow specialized for dementia care: **Triage → Research → Edit**. It performs web search, collects key facts from reputable medical sources, and generates a caregiver-friendly Markdown brief with references.

Research Process  **Report**

⌄ Report Outline

1. Executive Summary
2. Evidence Review
3. Practical Care Tips
4. Non-Pharmacologic Strategies First; then Medications
5. Red-Flags & When to Seek Care
6. Resources & Helplines

Word Count: 1081

# Non-Pharmacologic Approaches for Sleep Problems in Dementia

## Executive Summary

- **Sleep Disturbances Common**: Individuals with dementia often experience sleep problems, impacting both the individual and their caregiver.
- **Non-Pharmacologic Interventions**: Strategies like environmental adjustments and cognitive behavioral therapy have shown efficacy without the side effects associated with medications.
- **Practical Application**: Simple adjustments in daily routines, sleep hygiene, and creating a calming environment can significantly improve sleep quality.
- **Continual Assessment Needed**: Caregivers should monitor changes and adjust strategies accordingly, with a keen sense for potential red flags that require professional help.
- **Support Resources**: Various organizations offer guidance and support for caregivers dealing with sleep issues related to dementia.

## Evidence Review

Sleep disturbances are prevalent in individuals with dementia, including Alzheimer's disease, presenting as insomnia, nighttime awakenings, and daytime sleepiness. Research indicates that these disturbances can stem from neurobiological changes associated with dementia, altered circadian rhythms, and environmental factors. Non-pharmacologic strategies have gained support for managing sleep issues effectively:

1. **Cognitive Behavioral Therapy for Insomnia (CBT-I)**: Several studies highlight the effectiveness of CBT-I in improving sleep quality in older adults, including those with cognitive impairments.

    - According to the National Institute on Aging (NIA), CBT-I focuses on identifying negative thoughts about sleep and replacing them with positive ones.

2. **Environmental Modifications**: A systematic review found that optimizing the sleep environment (e.g., reduced noise levels, comfortable bedding, appropriate lighting) significantly aids in improving sleep quality.

    - Ensuring a cool, dark, and quiet room can facilitate better sleep patterns.

3. **Sleep Hygiene Education**: Education on maintaining good sleep hygiene is essential. Practices like consistent sleep schedules, avoidance of caffeine and heavy meals before bed, and establishing bedtime rituals can enhance sleep quality.

4. **Mindfulness and Relaxation Techniques**: Research shows that relaxation techniques can help manage anxiety and improve overall sleep quality. Techniques include meditation, deep breathing exercises, and gentle yoga.

5. **Diet and Exercise**: Diet influences sleep quality; for instance, foods rich in tryptophan can promote sleep. Regular physical activity also significantly contributes to improved sleep, although it's best to avoid vigorous exercises close to bedtime.

# Practical Care Tips

## Establishing Routines

- **Consistent Schedule**: Encourage adherence to regular sleep and wake times, even on weekends. Consistency helps regulate the body's internal clock.
- **Bedtime Rituals**: Create a calming pre-sleep routine, like reading, playing soft music, or engaging in gentle stretching activities. This signals the body that it's time to wind down.

## Modifying the Environment

- **Comfortable Sleeping Space**: Ensure the bedroom is quiet, dark, and cool. Consider using blackout curtains, earplugs, or white noise machines to reduce disruptive stimuli.
- **Comfortable Bedding**: Assess the comfort of bedding and pillows, ensuring they are adequate for the individual's preferences.

## Enhancing Communication

- **Clear Communication**: Use simple, clear communication and maintain a calm demeanor, especially during evening conversations. This helps reduce anxiety and confusion.
- **Emotional Reassurance**: Provide reassurance about sleep and encourage relaxation without pressure, creating a supportive atmosphere.

## Safety Precautions

- Ensure a clutter-free sleeping area to minimize the risk of falls during nighttime bathroom trips or when navigating the bedroom.

- Keep essential items within reach, including medications, glasses, and water.

# Non-Pharmacologic Strategies First; then Medications

## Non-Pharmacologic Approaches

- **Cognitive Behavioral Therapy for Insomnia (CBT-I)**

    - Focuses on modifying thoughts and behaviors associated with sleeping problems.
    - Effective for improving sleep in dementia patients without medications.

- **Light Therapy**: Exposure to bright light during the day can help regulate circadian rhythms and improve sleep at night.

- **Mindfulness Meditation**: Engaging in mindfulness can help alleviate anxiety and improve overall sleep quality.

## Medications (If Needed)

Individuals with dementia may sometimes require medications to address sleep issues. However, it's crucial to consider the following:

- **Potential Side Effects**: Many sleep medications can cause sedation, cognitive impairment, or exacerbate confusion—especially in older adults.
- **Collaborative Decision-Making**: Always discuss with healthcare professionals to weigh the benefits and risks before starting any medication.
- **Periodic Review**: If medications are prescribed, it's vital to regularly re-evaluate their effectiveness and side effects with a healthcare provider.

# Red-Flags & When to Seek Care

While many sleep disturbances can be managed at home, caregivers should be vigilant for the following red flags, indicating the need for professional intervention:

- **Acute Confusion or Delirium**: Sudden changes in behavior, confusion, or disorientation may be signs of an underlying health issue.
- **Injuries**: If the individual falls or gets hurt during attempts to get up at night, this requires immediate medical attention.
- **Dehydration**: Signs of dehydration, such as dry mouth or skin, may occur if an individual wakes frequently during the night.
- **New Neurological Deficits**: Any evidence of significant cognitive decline or new issues with memory, motor skills, or communication should prompt a visit to a healthcare provider.

# Resources & Helplines

- **National Institute on Aging (NIA):** [NIA Sleep and Aging](NIA Sleep and Aging) provides extensive resources on sleep issues in older adults.
- **Alzheimer's Association:** [Alzheimer's Association](Alzheimer's Association) offers numerous resources, including support groups and educational materials.
- **Local Community Services**: Many local agencies provide caregiver support services, respite care, and dementia-specific resources.

# References

- National Institute on Aging. (n.d.). Sleep and Aging. Retrieved from [https://www.nia.nih.gov](https://www.nia.nih.gov)

- Alzheimer's Association. (n.d.). Caregiving. Retrieved from [https://www.alz.org](https://www.alz.org)
- Mayo Clinic. (2022). Sleep Disorders in Dementia. Retrieved from [https://www.mayoclinic.org](https://www.mayoclinic.org)
- World Health Organization. (2020). Sleep Problems: Guidance on Sleep and Recovery. Retrieved from [https://www.who.int](https://www.who.int)
- American Academy of Sleep Medicine. (n.d.). Cognitive Behavioral Therapy for Insomnia. Retrieved from [https://aasm.org](https://aasm.org)

> References & Resources

Download Report

Start Alzheimer's Research

## ℹ️ Configuration

API Key is configured

🔄 Reset API Key

This tool provides AI-powered analysis of medical imaging using computer vision and radiology-style reasoning.

⚠️ DISCLAIMER: For **educational/informational** use only. Not medical advice. Do **not** make clinical decisions based solely on this output. Always consult qualified healthcare professionals.

# 🏥 Medical Imaging Diagnosis Agent

Upload a medical image for AI-assisted analysis (educational use only).

Upload Medical Image

Drag and drop file here
Limit 200MB per file • JPG, JPEG, PNG, DICOM, DCM

Browse files

3-alzheimers-brain-pasieka.jpg  216.1KB

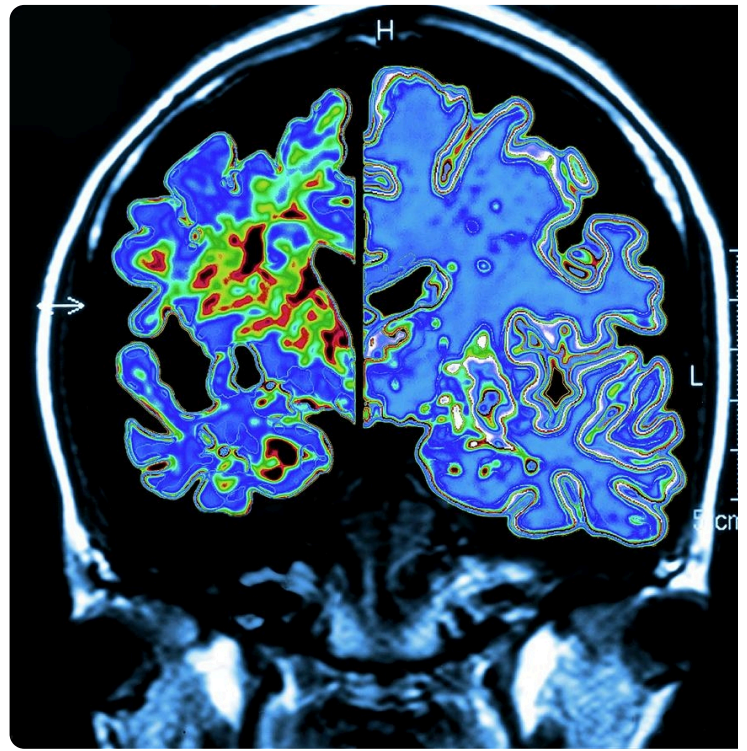

Uploaded Medical Image

🔍 Analyze Image

## 📋 Analysis Results

Okay, I will analyze the image you provided.

## 1) Image Type & Region

- **Modality:** Likely a coronal MRI or PET scan of the brain. The color overlay suggests it may be a functional image highlighting brain activity or metabolism.
- **Region:** Brain.
- **Quality:** The image quality is adequate for identifying major structural features.

## 2) Key Findings

- **Primary Observations:** There is a notable asymmetry between the left and right hemispheres. The left hemisphere appears to have reduced signal intensity (represented by blue/green/red color changes), which might indicate decreased activity or metabolism compared to the right hemisphere (predominantly blue).
- **Abnormalities/Artifacts:** The color-coded representation highlights regions of varying activity/metabolism. The asymmetry is the most prominent abnormality.
- **Severity:** The asymmetry appears moderate to severe, warranting further investigation.

## 3) Diagnostic Assessment (Non-diagnostic)

- The observed asymmetry could be related to various conditions, including:
    - **Neurodegenerative diseases** (e.g., Alzheimer's disease, frontotemporal dementia)
    - **Vascular events** (e.g., stroke)
    - **Focal brain lesions** (e.g., tumor)
- Given the asymmetry and possible involvement of the left hemisphere, language and motor functions should be assessed.
- **Critical findings:** The asymmetry should prompt urgent clinical attention and correlation with the patient's symptoms and medical history.

## 4) Patient-Friendly Explanation

This image is a scan of your brain. It looks like one side of your brain shows a different pattern compared to the other side. The colors represent how active different parts of your brain are. The differences observed could be related to several possible conditions. It's important to see a doctor for more tests to figure out what's going on.

## 5) Research Context

Here's a summary of key takeaways from reputable resources:

- **Brain Asymmetry:** Asymmetry in the brain can be associated with various conditions, including neurodevelopmental disorders, language disorders, and neurodegenerative diseases. (Wikipedia, NCBI)
- **Neuroimaging in Neurodegenerative Diseases:** MRI and PET scans are commonly used to identify patterns of brain atrophy or metabolic changes associated with neurodegenerative diseases. (Radiopaedia.org, PubMed Central)
- **Importance of Systematic Assessment:** When assessing brain MRI for neurodegenerative diseases, a careful and systematic approach is needed to identify pertinent features. (Radiopaedia.org)

**Important Note:** This analysis is based solely on the provided image and should not be considered a diagnosis. Clinical correlation with patient history and physical examination is essential. Please consult with a qualified healthcare professional for further evaluation.

---

Note: This analysis is AI-generated and must be reviewed by a qualified clinician.

## 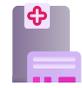 Medical Imaging Diagnosis Agent

Upload a medical image for AI-assisted analysis (educational use only).

Upload Medical Image

Drag and drop file here
Limit 200MB per file • JPG, JPEG, PNG, DICOM, DCM

Browse files

800wm.jpg  132.7KB

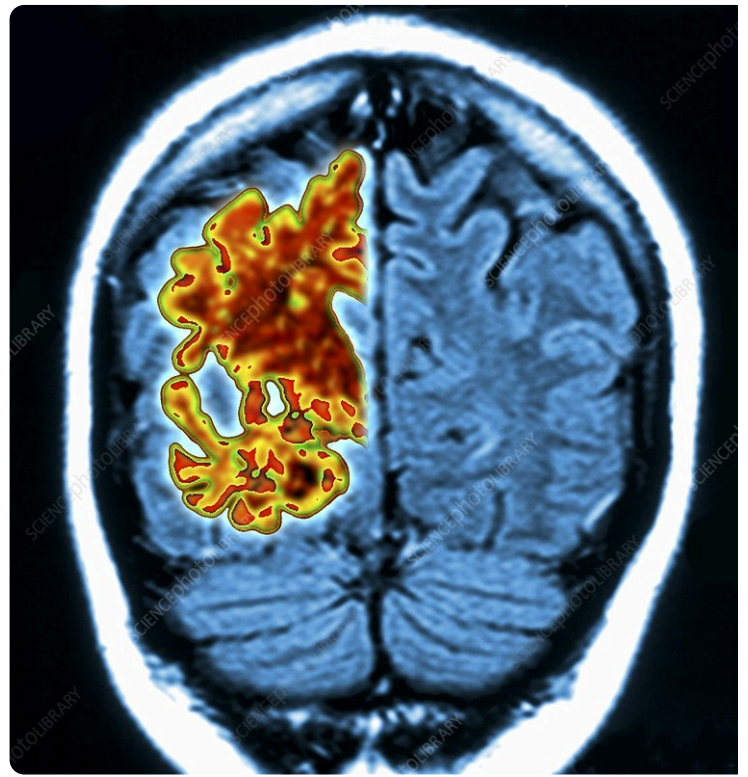

Uploaded Medical Image

🔍 Analyze Image

## 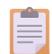 Analysis Results

---

Okay, I will analyze the provided medical image.

---

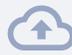

## 1) Image Type & Region

- **Modality:** Based on the characteristics of the image, it appears to be a coronal view of an MRI (Magnetic Resonance Imaging) scan of the brain. The coloring is not typical for an MRI, suggesting it might be a special sequence or has been processed for emphasis.
- **Region:** The image shows the brain.
- **Image Quality:** The image quality appears adequate for identifying major anatomical structures. However, the unusual coloration may hinder detailed assessment.

## 2) Key Findings

- **Primary Observations:** A large, well-defined area of abnormal signal intensity is present in the left hemisphere of the brain. It appears to affect a significant portion of the frontal lobe. The specific shape and location suggest involvement of multiple brain regions.
- **Abnormalities/Artifacts:** The highlighted area is very distinct and represents a significant abnormality.
- **Severity:** Qualitatively, the severity appears to be moderate to severe, given the size and location of the abnormality.

## 3) Diagnostic Assessment (Non-diagnostic)

- **Possible Considerations:** Given the location and appearance, possible considerations could include:
  - **Stroke (Infarction):** A large stroke could cause this pattern of damage.
  - **Tumor:** A slow-growing tumor might present similarly.
  - **Demyelinating Disease:** Although less typical, extensive demyelination could be considered.
  - **Encephalitis:** Certain infections can cause inflammation and damage of this extent.
- **Critical Findings:** The size and location of this abnormality warrant urgent clinical attention and further investigation to determine the underlying cause.

## 4) Patient-Friendly Explanation

This image is from an MRI scan of your brain. We see a large area on the left side that looks different from the rest. It's like a significant change in that part of your brain tissue. This could be due to several reasons, such as a blockage of blood flow (like a stroke), a growth, or inflammation. Because of what we see here, it's important to get more tests done quickly to figure out exactly what's going on and how best to address it.

## 5) Research Context

I will use the DuckDuckGo search tool to find recent medical literature or guidelines about similar patterns.

## Research Context: Large Frontal Lobe Lesions on MRI

Based on a quick search, here are some key takeaways regarding large frontal lobe lesions:

- **Differential Diagnosis:** A large lesion in the frontal lobe can have a broad differential diagnosis, including neoplastic, vascular, inflammatory, and infectious etiologies.
- **Glioblastoma Multiforme (GBM):** GBM is an aggressive primary brain tumor that can present as a sizable lesion in the frontal lobe.
- **Effects of Frontal Lobe Lesions:** Frontal lobe lesions can affect various cognitive functions, including executive functions, working memory, and attention.

**Reputable Resources:**

Given the limitations of the search, I would recommend exploring these resources for more information:

1. **Radiology Society of North America (RSNA):** Provides information on various imaging findings and conditions.
2. **National Institute of Neurological Disorders and Stroke (NINDS):** Offers information on stroke, brain tumors, and other neurological conditions.
3. **Mayo Clinic:** Provides patient-friendly information on a wide range of medical topics.

**Disclaimer:** This analysis is for informational purposes only and should not be considered a diagnosis or substitute for professional medical advice. A

qualified healthcare professional should evaluate the image and clinical findings to determine the appropriate course of action.

---

Note: This analysis is AI-generated and must be reviewed by a qualified clinician.

## Configuration

OpenAI API Key

[••••••••••••••••••••]  👁

OpenAI model  ⓘ

○ gpt-4o-mini
● gpt-4o

---

If you see OUTPUT_PARSING_FAILURE, this app will retry with a strict JSON fallback.

# 🕵️ Alzheimer's Web Scraping AI Agent

Scrape a web page with OpenAI + ScrapegraphAI. Includes a strict-JSON fallback if parsing fails.

## Target & Instructions

URL to scrape

```
https://my.clevelandclinic.org/health/diseases/9164-alzheimers-
```

What should the agent extract?

```
Can Alzheimer's be treated? If yes, how?
```

## Structured Output (Fallback Schema)

JSON Schema (used only if SmartScraperGraph output can't be parsed)

```
{
  "type": "object",
  "properties": {
    "title": {
      "type": "string"
    },
    "author": {
      "type": "string"
```

[ Scrape ]

SmartScraperGraph error: Failed to scrape after 1 attempts: BrowserType.launch: Executable doesn't exist at

```
/root/.cache/ms-playwright/chromium_headless_shell-
1181/chrome-linux/headless_shell
╔════════════════════════════════════════════════════════════╗
║ Looks like
Playwright was just installed or updated. ║ ║ Please run the
following command to download new browsers: ║ ║ ║ ║
playwright install ║ ║ ║ ║ <3 Playwright Team ║
╚════════════════════════════════════════════════════════════╝
```

Fallback completed.

## Strict JSON Fallback Output

```
{
  "summary" :
  "Alzheimer's disease is a neurodegenerative
  condition that affects memory, thinking, and
  other cognitive functions. While there is no
  cure, treatments can manage symptoms and slow
  progression. Medications like cholinesterase
  inhibitors, NMDA antagonists, and monoclonal
  antibodies are used to manage symptoms and slow
  the disease's progression. Clinical trials may
  offer access to new treatments."
    "key_points" : [
      0 :
      "Alzheimer's disease is the most common cause
      of dementia."
      1 :
      "There is no cure for Alzheimer's, but
      treatments can manage symptoms."
      2 :
      "Cholinesterase inhibitors and NMDA
      antagonists are common treatments."
      3 :
      "Monoclonal antibodies like lecanemab and
      donanemab target amyloid proteins."
```

```
            4:
            "Clinical trials may provide access to new
            treatments."
        ]
}
```

Tips if results are incomplete: • Make the prompt VERY explicit (list exact keys you need). • Simplify the schema to just the fields you must have. • Some pages are dynamic/authenticated; SmartScraper may need Playwright. If so, install with: %pip install -q 'scrapegraphai[playwright]' !python -m playwright install chromium

## Configuration

OpenAI API Key

[••••••••••••••••••]  👁

OpenAI model  ⓘ

○ gpt-4o-mini
● gpt-4o

---

If you see OUTPUT_PARSING_FAILURE, this app will retry with a strict JSON fallback.

# 🕵️ Alzheimer's Web Scraping AI Agent

Scrape a web page with OpenAI + ScrapegraphAI. Includes a strict-JSON fallback if parsing fails.

## Target & Instructions

URL to scrape

```
https://www.mayoclinic.org/diseases-conditions/alzheimers-dis
```

What should the agent extract?

```
Tell me the causations of Alzheimer's disease.
```

## Structured Output (Fallback Schema)

JSON Schema (used only if SmartScraperGraph output can't be parsed)

```
{
  "type": "object",
  "properties": {
    "title": {
      "type": "string"
    },
    "author": {
      "type": "string"
```

[ Scrape ]

SmartScraperGraph error: Failed to scrape after 1 attempts: BrowserType.launch: Executable doesn't exist at

```
/root/.cache/ms-playwright/chromium_headless_shell-
1181/chrome-linux/headless_shell
╔══════════════════════════════════════════════════════╗
║ Looks like Playwright was just installed or updated. ║
║ Please run the following command to download new browsers: ║
║                                                      ║
║     playwright install                               ║
║                                                      ║
║ <3 Playwright Team                                   ║
╚══════════════════════════════════════════════════════╝
```

Fallback completed.

## Strict JSON Fallback Output

```json
{
  "summary": "Alzheimer's disease is caused by a combination of genetic, lifestyle, and environmental factors that affect the brain over time. The disease is characterized by the formation of amyloid plaques and neurofibrillary tangles in the brain, which disrupt the function of neurons, leading to their damage and death. In less than 1% of cases, specific genetic changes almost guarantee the development of the disease. The damage often starts in the memory region of the brain and spreads in a predictable pattern, eventually causing the brain to shrink.",
  "key_points": [
    "Alzheimer's disease involves amyloid plaques and tau protein tangles in the brain.",
    "The exact causes are not fully understood but involve genetic, lifestyle, and environmental factors.",
    "In less than 1% of cases, specific genetic changes are responsible."
```

```
        3 :
        "The disease begins years before symptoms
        appear, starting in the memory region of the
        brain."
        4 :
        "By the late stage, the brain has
        significantly shrunk."
    ]
}
```

Tips if results are incomplete: • Make the prompt VERY explicit (list exact keys you need). • Simplify the schema to just the fields you must have. • Some pages are dynamic/authenticated; SmartScraper may need Playwright. If so, install with: %pip install -q 'scrapegraphai[playwright]' !python -m playwright install chromium



\end{document}